%% file: main.tex
\documentclass[conference]{IEEEtran}

\IEEEoverridecommandlockouts
\usepackage{cite}
\usepackage{amsmath,amssymb,amsfonts}
\usepackage{algorithm}
\usepackage{natbib}
\usepackage{amsmath}
\usepackage{xspace}

\usepackage{hyperref}

\usepackage{algpseudocode}
\usepackage{graphicx}
\usepackage{multirow}
\usepackage{booktabs}     
\usepackage[table]{xcolor} 
\newcommand{\blue}[1]{\textcolor{black}{#1}}

\newcommand{\maxtokens}{\textsf{max-output-tokens }}

\usepackage{textcomp}
\usepackage{xcolor}
\usepackage{subfigure}

\newcommand{\dummy}{\textsf{dummy-answer }}
\newcommand{\LMR}{LM$^{r}$\xspace}
\newcommand{\LMV}{LM$^{v}$\xspace}

\usepackage{makecell}
\def\BibTeX{{\rm B\kern-.05em{\sc i\kern-.025em b}\kern-.08em
    T\kern-.1667em\lower.7ex\hbox{E}\kern-.125emX}}
\begin{document}

\title{Membership Inference Attacks for Retrieval Based In-Context Learning for Document Question Answering 
}

\author{
    \IEEEauthorblockN{Tejas Kulkarni}
    \IEEEauthorblockA{Nokia Bell Labs\\    
    }
    \and
    \IEEEauthorblockN{Antti Koskela}
    \IEEEauthorblockA{Nokia Bell Labs}
    \and
    \IEEEauthorblockN{Laith Zumot}
    \IEEEauthorblockA{Nokia}
    
}
\maketitle

\begin{abstract}

We show that remotely hosted applications employing in‑context learning when augmented with a retrieval function to select in‑context examples can be vulnerable to membership inference attacks even when the service provider and users are separate parties.
\par We propose two black‑box membership‑inference attacks that exploit query text prefixes to distinguish member from non‑member inputs. The first attack uses a reference model to estimate an otherwise unavailable loss metric. The second attack improves upon it by eliminating the reference model and instead computing a membership statistic through a simple but novel weighted‑averaging scheme.  Our comprehensive empirical evaluations consider a stricter case in which the adversary has a paraphrased version of the text in the queries and show that our attacks can exhibit stronger resilience to paraphrasing and outperform three prior attacks in many cases with small number of prefixes. We also adapt an existing ensemble prompting defense to our setting, demonstrating that it substantially mitigates the privacy leakage caused by our second attack.

\end{abstract}

\input{introduction}

\input{problem_setting}

\input{attacks}

\input{attack_algorithms}

\input{experiments}

\input{defenses}

\input{limitations}

\bibliography{iclbib}
\appendix

\input{baselines}

\input{appendix}

\end{document}

%% file: introduction.tex
\section{Introduction}
\label{Sec:Introduction}
In-context learning (ICL)~\citep{iclpaper} is a popular prompt-engineering technique to enhance a generic language model’s response to a specific context/domain. A typical k-shot ICL pipeline prepares a prompt by concatenating  k task specific demonstration exemplars to the user query and then asks a language model (LM) to generate a response for an unknown query, conditioned on the demonstrations provided. The main attractive feature of ICL is that it does not involve compute heavy operations of updating model weights and typically API or prompt-only access to LM is sufficient.
\par ICL also shows a potential of becoming a two-party API service, where a service provider with high-quality task-specific demonstration examples, exposes API services (e.g. summarization~\cite{iclapplication1}, translation, question answering~\cite{iclapplication2}) to clients, enabling them to obtain more accurate responses to queries than those provided by generic LMs. 
\par Membership Inference Attacks~\cite{mia1,mia2} (MIAs) pose a challenge to adoption of LMs in privacy-critical use cases. An attacker conducting a  MIA holds a collection of test queries. Working within a defined threat model, the attacker interacts with a victim model to determine if a query was a part of the model's training pipeline. This seemingly benign binary information could have implications on individuals. For example, discovering that clinical notes about an individual's diagnosis are included in a medical question answering  tool tailored for a specific disease could disclose the patient's medical condition.  Additionally, a successful membership inference could be a starting point of a more damaging data extraction attack. 
Beyond assessing privacy leakage, MIAs are also used as an heuristic metric to evaluate success of  machine unlearning~\cite{mia_mul} methods.  
\par LMs are known to memorize the training datasets, and  privacy leakage in LLMs through MIAs is well studied~\citep{memorization1,miallm1,miallm2,miallm3,miallm4,miallm5}. Although the potential for training data leakage in LMs is real, exploiting memorization of training dataset is difficult, the MIAs on generic pretrained models could perform close to random guessing~\cite{miallm6}.
\par Compared to training data which is static and embedded into weights, information in an LM's context window could be more vulnerable and easier to uncover. During inference, the signals of recent prior exposure and lack of it could be more clearly separable. To this end, \cite{iclmia1,iclmia2,iclmia3} proposed MIAs that leverage few-shot ICL to detect the membership of a test data point in a prompt.

\par In this work, we consider a setting in which a document question answering (DQA) application using in-context learning is hosted on a remote server and demonstration examples chosen by server are hidden from user, and a malicious user is interested in inferring presence of his query example in the demonstration examples. From this perspective, we find the following shortcomings in the existing MIA works for ICL.
\begin{itemize}
\item \textbf{Limited to text classification:} Current MIA's on ICL focus only on classification tasks where a model chooses a label from a predefined list. While some of those can be adapted to generative tasks, we are not aware of any work that specifically targets a more complex task such as DQA which requires extracting information from input document to produce cohesive answers.  
    \item \textbf{Randomly sampled demonstrations:} The k-shot demonstrations in existing MIAs are selected randomly.  Picking examples unrelated to the test query is far from  practice~\cite{iclapplication1}, and could odd with the accuracy requirements of a service. It has been well-documented~\citep{lu2022fantastically,icl_bad_examples2,icl_bad_examples3} that the ICL outputs are sensitive to the demonstrations used, with randomly sampled demonstrations potentially leading to performance similar or worse than 0-shot predictions, undermining the purpose of guiding  examples. Therefore, exemplar selection is as an important research direction~\cite{icl_survey} in ICL. 
    
    Random demonstrations also reduce the practicality of the existing attacks because the chance of query entering into a prompt as a demonstration is quite slim~\footnote{In a $k$-shot prompt with a demonstration pool of size N, probability of sampling  the query (without replacement) is $\frac{k}{N }(1- \frac{1}{N})^{k-1}$.}. In other words, ICL with randomly sampled demonstrations is just less vulnerable to MIAs.     

    \item \textbf{Strong assumptions.} Attack proposed by \cite{iclmia1,iclmia2} rely on solely logit information which is often not available with the attacker when an ICL service is remotely hosted. 
    Let's consider two of the best performing attacks from \cite{iclmia3}, repeat and brainwash. 
    \par The Repeat attack takes a small prefix of the original query text, and asks the LM to predict the remaining part, which could be arbitrarily large. The experiments from \cite{iclmia3} only use the first 3 words of a query as the prefix. The embedding distance between true and predicted query suffix is used as the score. In practice, service providers could limit the number of output tokens generated for cost and security reasons. 
    \par The brainwash attack runs for a fixed number of iterations, attaching increasing number of copies of the original query paired with a wrong label in each iteration, with the intent of eventually forcing model to output the wrong label to input query. The attack uses the number of iterations needed to \emph{brainwash} a model as the membership signal, but it is unsuitable for generative tasks where query texts are typically long, because repeatedly adding query copies can overflow the model’s context window.

\end{itemize}
   
In summary, we find that all prior attacks are difficult to conduct, and possibly easier to detect and block. The question we aim to answer in this work is:

\vspace{0.2cm} 
\emph{Can we design effective membership inference attacks against ICL performing DQA task that only leverage model predictions?}

To answer this question affirmatively, it is reasonable to adopt a more realistic scenario where to meet its accuracy goals, the service provider uses a retrieval function to select prompt examples. A promising approach proposed by \cite{icl_index} is to pick demonstrations that are semantically similar to the text query itself. Note that embedding based $k$-nearest neighbor (kNN) search index is a standard component in information retrieval systems such as those designed for retrieval augmented generation (RAG)~\cite{icl_index}) and can be easily plugged into an existing ICL pipeline. A similar approach was taken by \cite{iclapplication1} that used ICL for medical dialogue summarization.
\par Interestingly, accuracy benefits of using nearest neighbors as in-context examples come at higher privacy risks: adversary's chances of inferring successful membership increase drastically because even a rephrased version of the query text has high likelihood appearing in the demonstration set, and also the prompt. We are not aware of any work that explored MIAs in this specific context despite its practical relevance and increased vulnerability.

\subsection{Our contributions}
\begin{itemize}
    \item In the setting where nearest neighbors are used as in-context examples, we propose two new attacks on ICL: The first attack employs a reference model to estimate a loss metric on the target model. The second attack functions only with the final predictions. Neither use accuracy/loss metrics internal to the victim model.
    \item  We focus our experiments on a key ICL use case: document question answering. To the best of our knowledge, this is the first work to instantiate MIAs for generative in‑context learning tasks as prior works have solely focused on in‑context learning for text classification~\footnote{\blue{In principle, our attacks can also be extended to text classification and summarization problems as well.}}.

    \item  We adapt prior attacks on classification to our setting and use them as baselines. Our experiments show that the proposed new attacks achieve higher TPR@low FPR in many cases, despite operating under much weaker assumptions. The higher TPR@low FPR values across datasets underscore the vulnerability of ICL when semantically similar demonstrations are used.

\end{itemize}

%% file: problem_setting.tex
\section{Problem Setting}
\subsection{In-context learning}
We have a server holding a private demonstration dataset $D =(x_1,x_2,\cdots,x_N)  \in \mathcal{X}^N$ of $N$ records to support document question answering task (DQA).  Each $x_i \in [N]$ is a triplet of the form $\langle t,q,a \rangle$, with $t$ representing a text paragraph and q and a forming a related question‑answer pair.

\par The service provider has converted the text part from each example in $D$ to embeddings and can retrieve the $k$-nearest neighbors examples from $D$ for a given text in a query using a suitable similarity search service $R_k$, such as a vector database. For simplicity, we assume that each document is converted into a single vector. The server also offers an interface to users to share their text query 
consisting of the text $t$ and the question $q$. 
\par  The user specifies $k\geq 1$, the number of in-context demonstrations needed. Upon receiving $\mathcal{Q}$ and $k$, server executes $R_k(D,t)$ using query text $t$ to retrieve kNNs. These neighbors are concatenated together in a single string using task appropriate demarcator words such as 'text:', 'question:', and 'answer:'.  
Server finally creates a single prompt composed of a task specific instruction, in-context demonstrations and string formatted user query. This process is described in Algorithm~\ref{alg:promptQA} in Appendix. Finally, the server uses the prompt to obtain predictions $P$ from an LM, which has a sufficiently large context window to process the prompt.

The sign '+' in Algorithm~\ref{alg:promptQA} represents the concatenation operation. The variable $Q$ signifies the user specific part of the prompt formed by combining text and a related question. The final word 'answer:' in $Q$ conveys our expectation to the  LM, indicating that it should generate a prediction after this word. The $k$ in-context demonstrations are used to learn the task specific input output mappings and improve over 0-shot prediction $\mathrm{P} = \mathrm{LM}\big(\mathrm{I} + Q\big)$, where the symbol $\mathrm{I}$ denotes the task specific instruction.
\subsection{Threat Model}
We assume that users trust the server with their queries. They are aware that the service uses examples with text semantically similar to the query text. Users have API‑only access to the LM and cannot view its internal workings, tokenizer, performance/loss metrics, similarity‑search function $R_k$, or modify the prompt instructions. Users can submit only a single query per API call and can set the maximum number of output tokens for that call; however, the actual input and output limits are enforced by the server. This means an attack such as \emph{Repeat} (Algorithm~\ref{alg:repeat}), which asks the LLM to predict a large number of tokens in a single API access given a short prefix, may have limited success.

 We further assume that client and  server interact only once for a single query, and server can retain previous interactions with the same client only for a limited time.
 \subsection{Adversary's goal} 
 The adversary has a test set with ground truth answers, and he aims to access the ICL service and determine, with a reasonable level of certainty, whether each test point in the set was part of the in-context examples used to generate the response.
 In other words, the adversary needs a scoring mechanism to distinguish between the following two cases: 
\begin{align*}
\mathrm{P} &= \mathrm{LM}\big(\mathrm{BuildPrompt}(D \cup \mathcal{Q},Q,t,k)\big), \\
\mathrm{P'} &= \mathrm{LM}\big(\mathrm{BuildPrompt}(D ,Q,t,k)\big),
    \end{align*}
where $\mathcal{Q} = \langle t,q,a \rangle$ and $Q$ is the concatenation $Q$ ='text:' + $t$ + 'question:' + $q$  +'answer:'.


%% file: attacks.tex
\begin{table}[!ht]
\centering
    \caption{Symbols used}

\begin{tabular}{ll}
\hline
Symbol & Description \\ \hline
I & task instruction \\
$\mathcal{I}$ & Prefix index set, $\mathcal{I} \subseteq [n]$ \\
$\mathcal{Q}$ & test query in a triplet or tuple format e.g., $\langle t,q,a\rangle$ or $\langle t,q,?\rangle$ \\
$P^v_i, P^r_i$ & ith element of the sets $P^v, P^r$ \\
$R_k$ & kNN retriever \\
$m$ & maximum number of output tokens \\  
$V$ & size of vocabulary in Algorithm~\ref{alg:logit} \\ 
$T$ & maximum number of tries in  Algorithm~\ref{alg:brainwash} \\
$p$ & fraction of text used as prefix \\
$t_{:i}$ & ith prefix of text $t$ \\
Q & test query concatenated as a formatted string \\
\LMV & Target (or victim) server language model \\
\LMR & Reference language model \\
D$^v$ or $D$ & Demonstration set for target server model \\
D$^{r}$ & Demonstration set for reference model \\
$\ell$ & membership score \\
\hline
\end{tabular}
    \label{tab:symbols}
\end{table}

%

\section{Proposed Attacks}
    \subsection{Attack 1: Using a reference model}
    \label{sec:refmodel}
    The Adversary does not have access to the loss values of predictions from the target model; however, it may be possible to estimate those values in some cases by locally mimicking the target service’s setup and leveraging ICL. We assume the attacker would like to predict the token‑wise conditional log‑probabilities and use their mean to differentiate between member and non‑member cases. To this end, the adversary locally runs a second model (the reference model) with a sufficiently large context window. The adversary can access the tokenizer and loss metrics of this model, which has not been trained on the test dataset in question.

    \par In this Section, we denote the target (or victim) server and reference models by \LMV and \LMR, with their respective demonstration sets denoted by $D^v$ and $D^r$. The adversary's $D^r$ consists of test queries, and optionally other demonstrations from outside the test set. The adversary knows the server uses kNNs of query text as context examples. In response, he creates a similar search mechanism  using $D^r$ to retrieve kNNs for a given query.

    For the text part $t$ of a query $\mathcal{Q}$ with $n$ tokens, an important user input to our attack is the ordered prefix index set $\mathcal{I} \cup \{n\} \subseteq [n]$. We denote as  $t_{:i}$ the ith prefix of text $t$, which consists only of the first $i$ words concatenated with spaces. The attacker carries out the following steps. This procedure is described more formally in Algorithm~\ref{alg:ref_model}.
    
    
    \begin{enumerate}
        \item For each  $ i \in \mathcal{I}$, construct a formatted test string $Q$ = 'text:' + $t_{:i}$, and obtain a single token prediction using \LMV and \LMR. These predictions aim to predict the $i+1$th token in the text space of the query text $t$. For the nth prefix however, i.e., the entire text, $Q$ is formed by 'text:' + $t$ + 'question' + $q$ + 'answer:'. We similarly  obtain predictions on both models. This prediction answers the question $q$. For all predictions, both the server and the adversary use their kNN indices to retrieve the in‑context examples. Let $P^r$ and $P^v$ be the ordered set containing the prefix predictions from \LMR and \LMV. Note that this is the only step that requires API calls to the server. The next steps are executed at the adversary's end.
    \item Compute the pairwise embedding distance for each pair of true word and its prediction in the text space, and populate the ordered set $S^r$, i.e. $S^{r} = \{ f(P^{r}_i,t_i)  \} ,\forall i \in \mathcal{I}$. The function $f$ transforms each word pair into embeddings, and then computes their dot product. 
    \item Compute the conditional log probabilities  $\log(\Pr\big[t_{i+1}|t_{:i}\big])$ for the true tokens on \LMR  for each prefix $\forall i  \in \mathcal{I} \cup \{n\}$ and populate them in the ordered set $L^{r}$. Note that the adversary does not have access these probabilities from \LMV. 
    
    \item Fit a 1-d kNN regression model M to predict $L^r$ from $S^r$. 
    \item Similar to step 2, obtain a dot product set $S^{v}$ using predictions from \LMV.
    \item Infer conditional log probabilities for \LMV using M, i.e. $\hat{L}^v = M(S^v)$. 
    \item Compute the mean $\ell = \frac{1}{|\hat{L}^{v}|} \sum_{i \in \mathcal{I} \cup \{n\}} \hat{L}_i^{v}$ as an estimate for the mean negative log-likelihood for the predictions, and use it as the membership signal. 
    \end{enumerate}
    
    \par\textbf{Intuition.}
        Figure~\ref{fig:attack1} in Appendix depicts the flow of Attack 1. The attacker has two sets of measurements obtained on \LMR, each of size $|\mathcal{I}|+1$. The first set $P^r$ contains the log probabilities for the $i+1$th token conditioned on all prior tokens. The second set $L^r$ comprises of the dot products (in the range $[-1,1]$) between the embeddings of the predictions and ground truth words reflecting their semantic similarities. 
    
    In case of membership, kNNs for even smaller prefixes could retrieve the same query with the ground truth answer as an in-context example. this could have the following consequences: (a) larger values in $L^r$ due to fewer mistakes in the next token predictions (b) the model assigning higher log probability values to the correct tokens. Conversely, the non-member queries could exhibit the opposite effect. The values corresponding to the smaller prefixes in $L^r$ and $S^r$ could be correlated, and the attacker can fit a kNN regression using these signals from \LMR to predict the log probabilities $\hat{L}^v$ originally unavailable for \LMV. The distributions of estimated values of $\ell$ for members might overlap, but we can expect the member values to be generally lower. A 1-d binary classifier can be trained using the score to learn the decision boundary. 

\par Appendix Figure~\ref{fig:qa_nll} seeks to test this hypothesis by comparing the distributions of $\ell$ values across three reference models for three DQA datasets. We also plot the true mean negative-likelihood distributions obtained directly from \LMV in the respective columns. We included all prefixes for this computation, i.e., $\mathcal{I}=[n]$ and use $k=1$. The main observation across the board is that $\ell$'s for the member queries compared to non-member queries are more spread out and have lower medians. Additionally, in many cases, member scores also take values in the ranges where there are no non-member scores. We also note that the gap between the two distributions widens as the reference model size increases. 

As we will show in the experiments, the reference model method enjoys the strongest accuracy gains. However, there are still limitations of this method. The adversary is burdened with the choice of the reference model and the cost of running it. Figure~\ref{fig:qa_nll} shows that two distributions are most separated  and member $\ell$'s take smaller values when \LMR and \LMV have comparable sizes. \blue{Moreover, this attacks relies on the assumption that the embedding relation of the tokens in the reference model are similar to that of the victim's LM. This may not hold true.} Finally, the attacker needs a regression model by predicting query text and answers using two formats ($Q$ = 'text' + $t_{:i}$ and $Q$ = 'text' + $t_{:i}$ + 'question:' + $q$ + 'answer:'), implying the service provider concatenates user supplied input string without verification and formatting to the rest of the prompt, leaving the user to decide the format. While this is still a weaker assumption compared to prior works, it is crucial to also think about scenarios when service provider would like remain in-charge of input formatting. We observed that effectiveness of the current attack reduces a lot when the adversary trains the regression model only using the answer predictions, and answers are short.  Our next attack fix both limitations, it only uses the answer predictions for membership inference, and does not employ an additional model.

\subsection{Attack 2: Using only the final predictions}
\label{sec:label_only}
Our second attack is based on the premise that a language model trained to follow instructions will attempt to complete the task even with incomplete information and may provide accurate answers if relevant information is present in the context provided. 

This Section onward, we refer the service provider's language model and demonstration set by LM and $D$. Using the prefix set $\mathcal{I}$, attack works as follows for a DQA task. 

\begin{enumerate}
    \item For each index $ i \in \mathcal{I}$, the attacker shares the query $\mathcal{Q} = \langle t_{:i},q,? \rangle$  with the service provider and obtains the prediction set $P^v$. Service provider uses kNNs for each text prefix $t_{:i}$ as in-context examples. Each element of $P^v$ is an answer to the question $q$ asked with the text prefix $t_{:i}$.  
    \item Using prediction set $P^{v}$, the attacker computes the score, $\ell = \sum_{i \in \mathcal{I} }  
 f(P_i^{v},a) \cdot \phi(i),$ and uses it for membership decision. The function $f$ is a task dependent distance function. For DQA tasks, $f$ could convert the predicted response $P^{v}_i$ and ground truth answer $a$ to embedding and compute their dot product~\footnote{For a classification problem, $f$ could simply return a binary value $\mathbb{I}_{P^{v}_i=a}$.}.  The function $\phi$ is a penalty function.  
\end{enumerate}

\textbf{Intuition.} The flow of Attack 2 is shown in Appendix Figure~\ref{fig:attack2}. The function $\phi$ is the most interesting part of the attack. Irrespective of membership status, the quality of target model's response is less likely to change for large enough prefix indices. For smaller prefixes however, and target model could behave differently in member and non-member case. In case of membership, model could still approximately answer due its access to the entire text in the context.
In case of non-membership, in the worst case, a model trained to answer truthfully could reply ``I do not know the answer`` or ``no information given in the text`` due to insufficient information in $t_{:i}$. In other words, for smaller prefixes, a relevant answer ($f(P^{v}_i,a) > 0$) is a strong membership signal, and a response indicating lack of knowledge ($f(P^{v}_i,a) \leq  0$) is a non-membership signal. We would like our score function to amplify early signals and provide diminishing returns for the subsequent answers. This can be achieved by using a decaying function such as $\phi(i)= \frac{1}{i}$ or $\frac{1}{\log(i)}$.

Figure~\ref{fig:lo_scores} plots the score distributions for member and non-member queries when all prefixes are used. We note that member scores roughly tend to take higher values than the non-members. The larger scores implies higher embedding distances for smaller prefixes which are further boosted with higher weights, and vice versa.

Note that this attack is difficult to detect because it does not violate any obvious service rules and gives the attacker plausible deniability. A negative response such as ``I don’t know the answer`` is possible even for queries with legitimate purposes. To enhance the attack further, the attacker could rearrange the query text so that information relevant to the question is moved toward the end.

%% file: attack_algorithms.tex
\begin{algorithm*}
\caption{Membership inference using a reference model.}\label{alg:ref_model}
\begin{algorithmic}[1]

\State Initialize the ordered sets $P^{r}=P^{v}=S^r=S^v=\{\}$
\Function{Pred-prefix-log-loss}{$\mathcal{Q}$, $\mathcal{I}$}

\For{$i  \in \mathcal{I} \cup \{n\}$}\Comment{Obtain prediction sets $P^r,P^v$ from reference and target model for each prefix index in  $\mathcal{I}$.}
\State $\langle t ,q,? \rangle=\mathcal{Q}$
\State $Q$ = 'text:' + $t_{:i}$
\State $m$=1
\If {$i = n$} \Comment{For the last token in the query, concatenate rest of the query'.}
\State $Q$ = $Q$ +  'question:' + $q$ + 'answer:' 
\State $m'=m$  
\EndIf
\State $P^{r} =P^{r} \cup \mathrm{LM}^{r} \big(\mathrm{BuildPrompt}(D^r,Q,t_{:i},k),\maxtokens=m\big)$
\State $P^{v} =P^{v} \cup \mathrm{LM}^{v} \big(\mathrm{BuildPrompt}(D^v,Q,t_{:i},k),\maxtokens=m\big)$
\EndFor
\State $L^{r} =L^{r} \cup \log(\Pr\big[t_{i+1}|t_{:i}\big]), \forall i  \in \mathcal{I}  \cup \{n\}$ \Comment{Compute the  negative log-likelihood for token prefixes on reference model.}
\State $S^{r}=\{f(P^{r}_i,t_i)\}, \forall  i \in \mathcal{I}  \cup \{n\} $ \Comment{Compute similarity  between each reference prediction and true query token using a scoring function $f$.} 
\State $S^{v}=\{f(P^{v}_i,t_i)\}, \forall  i \in \mathcal{I}  \cup \{n\}$ \Comment{Compute similarity  between each target prediction and true query token using a scoring function $f$.}
\State Train a kNN regression model $M$ to predict $L^r$ from $S^{r}$.

\State Infer the log losses for victim predictions using M, i.e., $\hat{L}^{v} = M(S^v)$
\State $\ell = \frac{1}{|\hat{L}^{v}|} \sum_{i \in \mathcal{I}  \cup \{n\}} \hat{L}_i^{v}$ \Comment{Compute predicted mean negative log-likelihood.}
\State \Return $\ell$
    \EndFunction

\end{algorithmic}
\end{algorithm*}

\begin{algorithm*}
\caption{Membership inference using only final predictions.}\label{alg:labelonly}
\begin{algorithmic}[1]
 
\State Initialize the ordered set $P^{v}=\{\}$
\Function{prediction-only}{$\mathcal{Q}$, $\mathcal{I}$}
\State $\langle t ,q,? \rangle=\mathcal{Q}$

\For{$i  \in \mathcal{I} $}\Comment{Obtain prediction sets $P^v$ from  target model for each prefix index in  $\mathcal{I}$.}
\State $Q$ = 'text:' + $t_{:i}$  +  'question:' + $q$ + 'answer:'

\State $P^{v} =P^{v} \cup \mathrm{LM}^{v} \big(\mathrm{BuildPrompt}(D,Q,t_{:i},k),\maxtokens=m)$
\EndFor

\State \Return $\ell = \sum_{i \in \mathcal{I}}  
 f(P_i^{v},a) \cdot \phi(i) $ \Comment{Compute weighted similarities between predicted and the true words.}
    \EndFunction

\end{algorithmic}
\end{algorithm*}

%% file: experiments.tex
\section{Experiments}

\begin{table}[ht]
\centering
\caption{DQA datasets used and their sizes.}
\label{tab:datasets}
\begin{tabular}{|l|c|c|}
\hline
\textbf{dataset} & \textbf{demonstration set} & \textbf{query set}  \\
\hline
SQuAD~\cite{squad}& $\sim$18.9k & 1k  \\
SQuADShifts~\cite{shifted_squad} & $\sim$4.6k & 1k  \\
NewsQA~\cite{newsqa} & $\sim$11k & 574     \\
\hline
\end{tabular}
\end{table}

\begin{figure}
    \centering
    \includegraphics[width=1\linewidth]{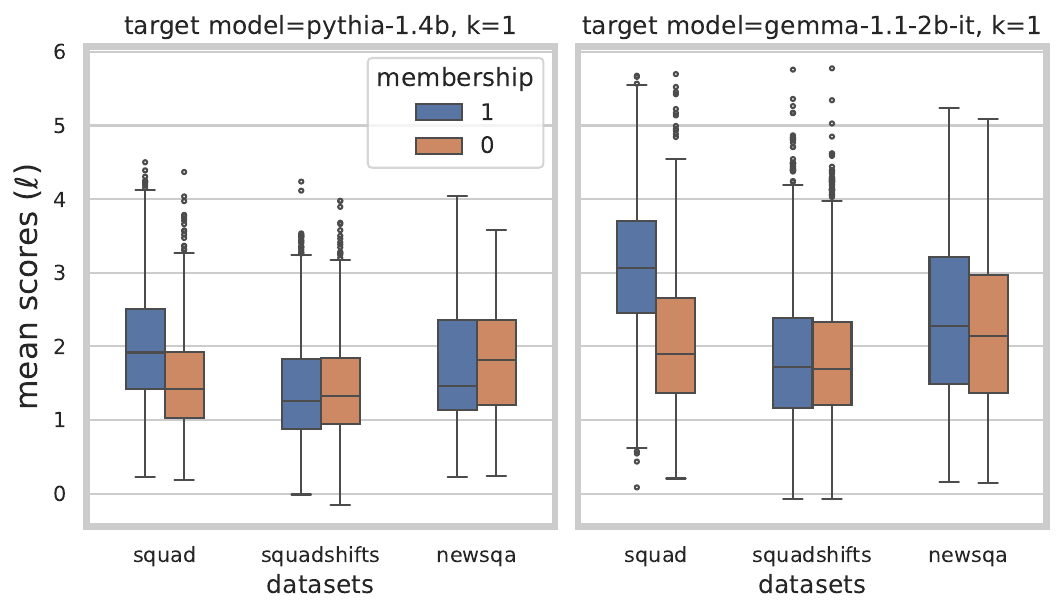}
    \caption{The distribution of mean membership scores for the attack from Section~\ref{sec:label_only} for member and non-member queries. }
    \label{fig:lo_scores}
\end{figure}

\subsection{Experimental Setup}
\par \textbf{Language Models.} We would like to support the DQA tasks through 1-shot ICL. We test our attacks mainly on two target models, Gemma-1.1-2B-IT~\cite{Gemma} and Pythia-1.4B~\cite{Pythia}. We also use Gemma-7B-IT~\cite{Gemma} in Section~\ref{sec:model_size} to validate the performance on a larger model. To account for the possibility that the parameter counts for target and reference model could differ a lot, our main experiments in Section~\ref{sec:main_results} use Pythia-160m~\cite{Pythia} as the reference model for the attack described in Section~\ref{sec:refmodel}. The use of Pythia suite was also motivated by our hardware constraints. While these models are just next token predictors, and not specifically designed for these tasks, they still provide acceptable performance. Additionally, Pythia's training dataset Pile~\cite{pile} is publicly documented, and possibly did not include the datasets we aim to use. This allows us to study the effect of membership inference in isolation. We select the next token in LMs using greedy decoding. To prevent truncation of the full query suffix, we set the maximum output tokens to 250 for the Repeat attack. For all other attacks, the maximum was set to 25.

\par \textbf{Datasets.} For DQA, we utilize the following benchmarking datasets accessed from Huggingface.  
\begin{itemize}
    \item \textbf{SQuAD v1.0:} crowd-sourced QA dataset based on Wikipedia articles. 
    \item \textbf{SQuADShifts:} a collection of four test datasets, encompassing the following domains: Wikipedia, New York Times articles and comments, Reddit articles and comments, and Amazon products and reviews.
    \item \textbf{NewsQA:} sourced from CNN news articles, answering questions requires understanding the concepts.  
    
\end{itemize}



We use the training split of each dataset as the demonstration set and the test/validation split as the query set. SQuADShifts does not have a train set. We combine the four small datasets together and create a random demonstration and a test split. To respect the model context window, we only consider demonstrations with text part smaller than 500 words. Some DQA datasets contain multiple questions for each dialog/paragraph. We randomly sample a single question-answer pair per text to de-duplicate the dataset. Our trimmed dataset sizes are reported in Table~\ref{tab:datasets}.

\textbf{Attack settings.} In the reference model attack, we use 5 neighbors for in the kNN regression model, while the second attack uses $\phi(i)=\frac{1}{i}$ as the penalty function.

\par \textbf{Similarity Search.} To build kNN indices, the text dialogues/paragraphs from each record are converted to embeddings using the all-MiniLM-L6-v2~\cite{embmodel} model which provides a good balance between the inference performance and embedding quality. It produces embeddings of unit length with dimension 384. We use FLAT indices for retrievals, which are constructed using FAISS~\cite{douze2024faiss} library.
\par \textbf{Evaluation Settings.} The service provider's kNN index differentiates abetween member and non-member scenarios. In the member case, the index uses both demonstration and query sets, allowing the query to be retrieved as an in-context example. In contrast, the non-member index is based solely on the demonstration set. The adversary's index in the reference model attack uses only the query set records, regardless of the case. 

\par In typical MIA assessments, target and query records are assumed to be identical. However, this assumption is unrealistic, as adversaries may have access to alternative versions of the same text. Paraphrased queries reduce the attack surface, and paraphrasing is commonly viewed as a defense against MIAs.  However, this is not true in our case. The use of similarity search does not significantly  degrade the attacker's likelihood of successfully inferring the membership status.    
To accommodate for this scenario, we use Llama-3-70B-IT~\cite{llama3} to paraphrase the query datasets by inverting voice (active to passive and vice versa), replacing words with synonyms. 

\par \textbf{Prefix sets.} Both attacks make '$n$' API calls to test the membership of a single query in the worst case, which could be costly. Therefore, our experiments also verify the attack performance when a much smaller prefix set is used instead. We include a column named '$p$' which shows metrics for the fraction of prefixes considered for running attack in our tabular results. For the second attack, $p=0.1$ means the attack used the first 10\% prefixes from the left. For the  attack using a reference model, we also include the nth prefix (i.e. the entire text) in the attack additionally.

\par \textbf{Membership inference methodology.} Typically MIAs use a threshold to distinguish between members and non-members. For complex tasks however, finding such a single threshold can be challenging for complex tasks due to overlapping score distributions (e.g. see Figure~\ref{fig:qa_nll}). We rely on the classical approach of training a binary classifier to predict membership across all methods instead. 
To create a balanced attack dataset, we execute each query both on member and non-member indices, and obtain two scores. After the dataset is prepared, we predict memberships with a 1-$d$ random forest classifier (20 estimators), evaluated under 10-fold stratified cross-validation.

\par \textbf{Evaluation Metrics.} Our main metric is TPR@low FPR~\cite{tpr_at_low_fpr}, which directly assesses an attack's ability to correctly flag members with low non-member accusations. For completeness, we also report Area Under the Curve (AUC) for ROC, providing a comprehensive view of separability across all possible thresholds.

\par \textbf{Implementation.} All attacks are implemented in PyTorch. We rely on Huggingface for dataset and model access, and scikit-learn for training membership inference models and calculating metrics.

\subsection{Baseline Methods} To the best of our knowledge, no prior work has focused on attacking ICL used for generative tasks. However, it is possible to extend methods from \cite{iclmia1,iclmia2,iclmia3} for the DQA task since these are agnostic to the fashion in which in-context examples are chosen. We compare our attack's performance against the following baseline methods, which are described in Appendix Algorithms~\ref{alg:logit}, \ref{alg:repeat}, and \ref{alg:brainwash}.

\begin{itemize}
    \item \textbf{Logit based~\cite{iclmia1,iclmia2}.}  For query $\mathcal{Q}$, the adversary obtains model prediction set $P^v$ and the logit matrix $L^v$. He accesses the logits values corresponding to the true answer a which is already knows. The adversary  uses the mean of these values as membership score. These can be sufficiently different for member and non-member queries.
    \item \textbf{Repeat~\cite{iclmia3}}.  The attack submits a small prefix of the query text to the server and asks to predict the suffix. The embedding distance between the original and predicted suffix is used as the membership score. In Algorithm~\ref{alg:repeat}, we denote by $p \in [0,1]$, the fraction of text used as a prefix. For our experiments, we set $p$ to 0.1.  
    \item \textbf{Brainwash~\cite{iclmia3}}. This iterative attack queries the ICL service by repeatedly prepending increasing number of copies of the test query with a dummy/incorrect answer to the original query. The number of iterations required (T in Algorithm~\ref{alg:brainwash})  to force the wrong response is treated as the membership score. For our simulation, we set $T=10$, a greater value than 5 that the authors used.

\end{itemize}

We stress that these baselines are  compared only for a perspective. As mentioned in Section~\ref{Sec:Introduction}, these cannot be applied in a two-party setup such as ours or unsuitable for generative tasks. For example, sending a large string containing several copies of the query to the server in a single API call in the brainwash attack could easily violate service's input token limit, making it difficult to carry out this attack effectively.

\section{Results}

\subsection{Main results}
\label{sec:main_results}
Tables~\ref{tab:newsqa_para}, \ref{tab:shifted_squad_para}, and \ref{tab:squad_para} report the performance of all methods for 1-shot ($k=1$) ICL when adversary has a paraphrased version of the text in each query. In each column, for our methods, we  highlight the metric corresponding to the smallest value of $p$ that outperformed all three baselines. The main high-level observation across the tables is that the attack using a reference model usually has the strongest AUC, and it surpasses the baselines just with 10\% of the query tokens in some cases. The second attack shows a steady performance increase with more prefixes and performs at least as good as, and often better than, the brainwash and logit attacks for newsqa and SQuADShifts datasets. The reference model attack does better than the prediction only attack because it captures more granular membership signals, i.e. token-level vs. sentence-level.
\par The repeat attack outperforms our methods on TPR @ low FPR metrics on the SQuAD dataset (Table~\ref{tab:squad_para}). However, unlike our method, it attempts to predict 90\% of the query text in one API call, which could be challenging in practice due to the risk of raising an anomaly alert or having the response truncated. 
\par The logit based method is probably the the most direct way to inference membership. Ignoring its practicality, one may expect it to emerge as the best method. However, this is not the case. This could be because unlike classification problems, the output tokens for generative tasks take values from a much larger set, and LM could output one of several variations of the same word. For example, LM could predict the answer $\$1$ as 'one dollar', '1 dollar', or 'one US dollar'. And the logit for the true answer $\$1$, could be lower than its variations, diluting the membership signal.  
\par The last observation is that for the brainwash method on the Pythia model provides very little membership signal compared to Gemma. While it is difficult to argue conclusively, the sensitivity of ICL to the placement of examples combined with Pythia's short context window could be to blame. We position the in-context example before the “brainwash” examples. This order could cause the model to forget the in-context example.  Based on the same premise, an alternative hypothesis is that after processing a cohesive text and its accompanying question, a model that is not explicitly trained to follow instructions could weigh tokens relevant to a logical answer more heavily than unnatural sounding dummy answer i.e., $\perp$. In both cases, model returns either a relevant answer or the dummy answer in the same iteration for both member and non-member cases, making it difficult to distinguish between them.

\subsection{Effects of varying the reference model}
For the results in Tables~\ref{tab:newsqa_para},\ref{tab:shifted_squad_para}, and \ref{tab:squad_para}, Pythia-160m was used as the reference model. Table~\ref{tab:vary_ref} checks the performance consistency of our attack against two more reference models. We highlight the largest metric for each value of $p$. The main high-level observation is that the optimal reference model is dataset and prefix length dependent. We also note that the  changes in AUC's across reference models are not dramatic in most cases. The largest variance in the TPR@ low FPR metrics are observed for the newsQA dataset which has much longer articles than other datasets. In many cases, it is possible to improve over Pythia-160m, however the trend is hard to establish.  The main takeaway from this Table is that you can outperform the baselines even with a suboptimal choice of reference model.   
\subsection{Effects of paraphrasing} Our prior experiments considered that the texts in the attacker's queries are rephrased. In the reference model attack, replacing exact prefixes with prefixes from paraphrased text may lower alignment between ground-truth words and model outputs, obscuring membership cues.  In the second attack, membership signals could only become apparent after processing longer prefixes of the paraphrased text, but these signals would be weakened by much larger inverse weights.

Therefore, an interesting study is to the check the extent to which paraphrasing affects the attack success rates. Appendix Tables~\ref{tab:shifted_squad_exact}, \ref{tab:squad_exact}, and \ref{tab:newsqa_exact}, show the success metrics for the counterfactual case when the adversary attacks with the exact version of the query texts. We compare the paired tables (\ref{tab:shifted_squad_para},\ref{tab:shifted_squad_exact}), (\ref{tab:squad_para},\ref{tab:squad_exact}), and (\ref{tab:newsqa_para},\ref{tab:newsqa_exact}) and confirm that the attack strengths indeed rise when the exact texts are used. A surprising outcome is that baselines almost always perform better than our attacks with the exact texts, but not with paraphrased texts. This highlights our attack's resilience against paraphrasing.

\subsection{Effect of model size}
\label{sec:model_size}
Table~\ref{tab:gamma7b} shows the attack performance on Gemma-7B-IT model.
When compared to Tables~\ref{tab:newsqa_para},~\ref{tab:squad_para}, and \ref{tab:shifted_squad_para}, the main observation is that the performance of the brainwash, repeat, and our second attack reduces compared to the smaller models. The reference model attack however performs similar to the logit attack.

\begin{table*}[htbp]
\caption{\textbf{newsQA}: The table compares cross-validated attack metrics, with the attacker having a paraphrased version the text in the test queries. In each column, we highlight the metrics corresponding to the value of '$p$' where our attack exceeds the baselines.}
\label{tab:newsqa_para}
\centering
\begin{tabular}{ll|r|r|r|r|r|r}
\toprule
target model & $k$ & attack  & $p$ & AUC & TPR@1\%FPR & TPR@5\%FPR & TPR@10\%FPR \\
\midrule
\multirow[t]{13}{*}{Gemma-1.1-2B-IT} & \multirow[t]{13}{*}{1} & brainwash & - & 0.743 & 0.091 & 0.203 & 0.310 \\
\cline{3-8}
 &  & repeat & - & 0.619 & 0.066 & 0.130 & 0.242 \\
\cline{3-8}
 &  & logit & - & 0.880 & 0.266 & 0.492 & 0.560 \\
\cline{3-8}
 &  & \multirow[t]{3}{*}{reference (our)} & 0.1 & \textbf{0.886} & 0.170 & \textbf{0.530} & \textbf{0.720} \\
 &  &  & 0.2 & 0.888 & 0.155 & 0.450 & 0.706 \\
 &  &  & 1 & 0.893 & 0.166 & 0.553 & 0.769 \\
\cline{3-8}
 &  & \multirow[t]{7}{*}{} & 0.05 & 0.717 & 0.181& 0.271 & 0.410 \\
 &  &  & 0.1 & \textbf{0.798} & 0.197 & 0.303 & 0.523 \\
 &  & predictions only (ours) & 0.2 & 0.851 & 0.216 & 0.352 & 0.542 \\
 &  &  & 0.3 & 0.864 & 0.216 & 0.355 & 0.555 \\
 &  &  & 0.5 & 0.878 & 0.245 & 0.365 & \textbf{0.587} \\
 &  &  & 0.8 & 0.886 & 0.252 & 0.374 & 0.610 \\
 &  &  & 1 & 0.887 & 0.265 & 0.381 & 0.619 \\
\bottomrule
\multirow[t]{13}{*}{Pythia-1.4B} & \multirow[t]{13}{*}{1} & brainwash & - & 0.517 & 0.011 & 0.049 & 0.108 \\
\cline{3-8}
 &  & repeat & - & 0.552 & 0.050 & 0.090 & 0.143 \\
\cline{3-8}
 &  & logit & - & 0.861 & 0.570 & 0.453 & 0.707 \\
\cline{3-8}
 &  & \multirow[t]{3}{*}{reference (our)} & 0.1 & \textbf{0.862} & 0.320 & \textbf{0.462} & \textbf{0.755} \\
 &  &  & 0.2 & 0.892 & 0.352 & 0.412 & 0.735 \\
 &  &  & 1 & 0.847 & 0.316 & 0.380 & 0.760 \\
\cline{3-8}
 &  & \multirow[t]{7}{*}{} & 0.05 & \textbf{0.707} & 0.160 & \textbf{0.243} & 0.430 \\
 &  & predictions only (ours)  & 0.1 & 0.783 & 0.210 & 0.330 & 0.507 \\
 &  &  & 0.2 & 0.826 & 0.263 & 0.403 & 0.590 \\
 &  &  & 0.3 & 0.840 & 0.287 & 0.457 & 0.593 \\
 &  &  & 0.5 & 0.855 & 0.337 & 0.470 & 0.620 \\
 &  &  & 0.8 & 0.862 & 0.370 & 0.497 & 0.630 \\
 &  &  & 1 & 0.864 & 0.363 & 0.507 & 0.640 \\
\bottomrule
\end{tabular}
\end{table*}

\begin{table*}[htbp]
\caption{\textbf{SQuADShifts}: The table compares cross-validated attack metrics, with the attacker having a paraphrased version the text in the test queries. In each column, we highlight the metrics corresponding to the value of '$p$' where our attack exceeds the baselines.}
\label{tab:shifted_squad_para}
\centering
\begin{tabular}{ll|r|r|r|r|r|r}
\toprule
target model & $k$ & attack  & $p$ & AUC & TPR@1\%FPR & TPR@5\%FPR & TPR@10\%FPR \\
\midrule
\multirow[t]{13}{*}{Gemma-1.1-2B-IT} & \multirow[t]{13}{*}{1} & brainwash& - & 0.736 & 0.112 & 0.310 & 0.405 \\
\cline{3-8}
 &  & repeat & - & 0.564 & 0.090 & 0.149 & 0.228 \\
\cline{3-8}
 &  & logit& - & 0.608 & 0.060 & 0.123 & 0.184 \\
\cline{3-8}
 &  & \multirow[t]{3}{*}{reference (our)} & 0.1 & \textbf{0.854} & \textbf{0.122} & \textbf{0.370} & \textbf{0.646} \\
 &  &  & 0.2 & 0.876 & 0.119 & 0.390 & 0.662 \\
 &  &  & 1 & 0.882 & 0.085 & 0.470 & 0.748 \\
\cline{3-8}
 &  & \multirow[t]{7}{*}{} & 0.05 & 0.597 & 0.058 & 0.168 & 0.252 \\
 &  & predictions only (ours)  & 0.1 & 0.660 & 0.117 & 0.221 & 0.313 \\
 &  &  & 0.2 & 0.720 & \textbf{0.129} & 0.273 & 0.377 \\
 &  &  & 0.3 & \textbf{0.746} & 0.133 & 0.294 & 0.405 \\
 &  &  & 0.5 & 0.768 & 0.146 & 0.301 & \textbf{0.441} \\
 &  &  & 0.8 & 0.781 & 0.153 & \textbf{0.316} & 0.474 \\
 &  &  & 1 & 0.785 & 0.155 & 0.318 & 0.484 \\
\bottomrule
\multirow[t]{13}{*}{Pythia-1.4B} & \multirow[t]{13}{*}{1} & brainwash& - & 0.502 & 0.010 & 0.050 & 0.100 \\
\cline{3-8}
 &  & repeat& - & 0.622 & 0.174 & 0.236 & 0.308 \\
\cline{3-8}
 &  & logit& - & 0.619 & 0.114 & 0.171 & 0.224 \\
\cline{3-8}
 &  & \multirow[t]{3}{*}{reference (our)} & 0.1 & \textbf{0.856} & 0.067 & \textbf{0.387} & \textbf{0.632} \\
 &  &  & 0.2 & 0.883 & \textbf{0.178} & 0.454 & 0.711 \\
 &  &  & 1 & 0.876 & 0.182 & 0.454 & 0.738 \\
\cline{3-8}
 &  & \multirow[t]{7}{*}{} & 0.05 & 0.568 & 0.022 & 0.087 & 0.161 \\
 &  &  & 0.1 & 0.605 & 0.072 & 0.147 & 0.238 \\
 &  & predictions only (ours)  & 0.2 & \textbf{0.648} & 0.086 & 0.179 & 0.281 \\
 &  &  & 0.3 & 0.670 & 0.092 & 0.200 & 0.297 \\
 &  &  & 0.5 & 0.692 & 0.105 & 0.221 & \textbf{0.312} \\
 &  &  & 0.8 & 0.706 & 0.106 & \textbf{0.238} & 0.336 \\
 &  &  & 1 & 0.711 & 0.107 & 0.249 & 0.352 \\
\bottomrule
\end{tabular}

\caption{\textbf{SQuAD}: The table compares cross-validated attack metrics, with the attacker having a paraphrased version the text in the test queries. In each column, we highlight the metrics corresponding to the value of '$p$' where our attack exceeds the baselines.}
\label{tab:squad_para}
\centering
\begin{tabular}{ll|r|r|r|r|r|r}
\toprule
target model & $k$ & attack  & $p$ & AUC & TPR@1\%FPR & TPR@5\%FPR & TPR@10\%FPR \\
\midrule
\multirow[t]{13}{*}{Gemma-1.1-2B-IT} & \multirow[t]{13}{*}{1} & brainwash& - & 0.723 & 0.123 & 0.257 & 0.352 \\
\cline{3-8}
 &  & repeat& - & 0.673 & 0.125 & 0.238 & 0.359 \\
\cline{3-8}
 &  & logit& - & 0.611 & 0.030 & 0.054 & 0.110 \\
\cline{3-8}
 &  & \multirow[t]{3}{*}{reference (our)} & 0.1 & \textbf{0.872} & \textbf{0.129} & \textbf{0.418} & \textbf{0.698} \\
 &  &  & 0.2 & 0.883 & 0.112 & 0.445 & 0.687 \\
 &  &  & 1 & 0.880 & 0.123 & 0.486 & 0.763 \\
\cline{3-8}
 &  & \multirow[t]{7}{*}{} & 0.05 & 0.571 & 0.035 & 0.104 & 0.164 \\
 &  & predictions only (ours)  & 0.1 & 0.648 & 0.039 & 0.140 & 0.248 \\
 &  &  & 0.2 & 0.721 & 0.076 & 0.201 & 0.324 \\
 &  &  & 0.3 & \textbf{0.752} & 0.123 & 0.220 & \textbf{0.360} \\
 &  &  & 0.5 & 0.775 & \textbf{0.130} & \textbf{0.261} & 0.388 \\
 &  &  & 0.8 & 0.785 & 0.128 & 0.274 & 0.412 \\
 &  &  & 1 & 0.787 & 0.122 & 0.280 & 0.427 \\
\bottomrule
\multirow[t]{13}{*}{Pythia-1.4B} & \multirow[t]{13}{*}{1} & brainwash & - & 0.502 & 0.013 & 0.051 & 0.101 \\
\cline{3-8}
 &  & repeat & - & 0.722 & 0.346 & 0.426 & 0.494 \\
\cline{3-8}
 &  & logit & - & 0.679 & 0.110 & 0.215 & 0.272 \\
\cline{3-8}
 &  & \multirow[t]{3}{*}{reference (our)} & 0.1 & \textbf{0.845} & 0.077 & 0.301 & \textbf{0.569} \\
 & &   & 0.2 & 0.883 & 0.117 & 0.422 & 0.694 \\
 &  &  & 1 & 0.875 & 0.181 & \textbf{0.456} & 0.710 \\
\cline{3-8}
 &  & \multirow[t]{7}{*}{} & 0.05 & 0.542 & 0.027 & 0.075 & 0.148 \\
 &  & predictions only (ours)  & 0.1 & 0.582 & 0.031 & 0.095 & 0.170 \\
 &  &  & 0.2 & 0.627 & 0.041 & 0.127 & 0.226 \\
 &  &  & 0.3 & 0.647 & 0.047 & 0.133 & 0.248 \\
 &  &  & 0.5 & 0.666 & 0.055 & 0.159 & 0.271 \\
 &  &  & 0.8 & 0.676 & 0.066 & 0.164 & 0.286 \\
 &  &  & 1 & 0.678 & 0.072 & 0.169 & 0.292 \\
\bottomrule

\end{tabular}

\end{table*}

\begin{table*}[htbp]    
\caption{The table compares the effect of varying the reference model. We report the cross-validated attack metrics, with the user having a paraphrased version the text in the test queries.  We highlight the largest metric for each value of $p$.}
\label{tab:vary_ref}
\centering
\begin{tabular}{r|r|rrrrrr}
\toprule
 dataset & target model & $p$ & reference& AUC & TPR@1\%FPR & TPR@5\%FPR & TPR@10\%FPR \\
\midrule
\multirow[t]{9}{*}{SQuAD} & \multirow[t]{9}{*}{Gemma-1.1-2B-IT} & \multirow[t]{3}{*}{0.1} & Gemma-1.1-2B-IT & 0.873 & 0.119 & \textbf{0.425} & \textbf{0.717} \\
 &  &  & Pythia-1.4B & \textbf{0.883} & 0.096 & 0.395 & 0.673 \\
 &  &  & Pythia-160m & 0.872 & \textbf{0.129} & 0.418 & 0.694 \\
\cline{3-8}
 &  & \multirow[t]{3}{*}{0.2} & Gemma-1.1-2B-IT & \textbf{0.875} & \textbf{0.152} & 0.426 & \textbf{0.719} \\
 &  &  & Pythia-1.4B & 0.874 & 0.107 & 0.438 & 0.695 \\
 &  &  & Pythia-160m & 0.883 & 0.112 & \textbf{0.445} & 0.687 \\
\cline{3-8}
 &  & \multirow[t]{3}{*}{1} & Gemma-1.1-2B-IT & \textbf{0.884} & 0.093 & 0.397 & 0.750 \\
 &  &  & Pythia-1.4B & 0.877 & 0.095 & 0.454 & 0.694 \\
 &  &  & Pythia-160m & 0.880 & \textbf{0.123} & \textbf{0.486} & \textbf{0.763} \\
\cline{2-8}
\multirow[t]{9}{*}{} & \multirow[t]{9}{*}{Pythia-1.4B} & \multirow[t]{3}{*}{0.1} & Gemma-1.1-2B-IT & \textbf{0.874} & \textbf{0.136} & \textbf{0.379} & \textbf{0.626} \\
 &  &  & Pythia-1.4B & 0.866 & 0.062 & 0.302 & 0.613 \\
 &  &  & Pythia-160m & 0.845 & 0.077 & 0.301 & 0.563 \\
\cline{3-8}
 &  & \multirow[t]{3}{*}{0.2} & Gemma-1.1-2B-IT & \textbf{0.875} & 0.092 & 0.381 & 0.694 \\
 &  &  & Pythia-1.4B & 0.872 & 0.106 & 0.396 & \textbf{0.707} \\
 &  &  & Pythia-160m & 0.877 & \textbf{0.117} & \textbf{0.422} & 0.694 \\
\cline{3-8}
 &  & \multirow[t]{3}{*}{1} & Gemma-1.1-2B-IT & 0.873 & 0.150 & 0.381 & 0.687 \\
 &  &  & Pythia-1.4B & \textbf{0.880} & 0.101 & 0.438 & 0.654 \\
 &  &  & Pythia-160m & 0.877 & \textbf{0.181} & \textbf{0.456} & \textbf{0.710} \\
\cline{1-8}
 \multirow[t]{9}{*}{SQuADShifts} & \multirow[t]{9}{*}{Gemma-1.1-2B-IT} & \multirow[t]{3}{*}{0.1} & Gemma-1.1-2B-IT & \textbf{0.901} & 0.156 & 0.449 & \textbf{0.774} \\
 &  &  & Pythia-1.4B & 0.892 & \textbf{0.160} & \textbf{0.525} & 0.758 \\
 &  &  & Pythia-160m & 0.854 & 0.122 & 0.370 & 0.646 \\
\cline{3-8}
 &  & \multirow[t]{3}{*}{0.2} & Gemma-1.1-2B-IT & \textbf{0.886} & \textbf{0.179} & \textbf{0.526} & \textbf{0.751} \\
 &  &  & Pythia-1.4B & 0.873 & 0.132 & 0.464 & 0.675 \\
 &  &  & Pythia-160m & 0.878 & 0.119 & 0.390 & 0.662 \\
\cline{3-8}
 &  & \multirow[t]{3}{*}{1} & Gemma-1.1-2B-IT & \textbf{0.901} & \textbf{0.169} & \textbf{0.589} & \textbf{0.782} \\
 &  &  & Pythia-1.4B & 0.890 & 0.136 & 0.466 & 0.765 \\
 &  &  & Pythia-160m & 0.882 & 0.085 & 0.470 & 0.748 \\
\cline{2-8}
\multirow[t]{9}{*}{} & \multirow[t]{9}{*}{Pythia-1.4B} & \multirow[t]{3}{*}{0.1} & Gemma-1.1-2B-IT & \textbf{0.908} & \textbf{0.157} & \textbf{0.528} & \textbf{0.812} \\
 &  &  & Pythia-1.4B & 0.884 & 0.080 & 0.367 & 0.710 \\
 &  &  & Pythia-160m & 0.856 & 0.067 & 0.387 & 0.632 \\
\cline{3-8}
 &  & \multirow[t]{3}{*}{0.2} & Gemma-1.1-2B-IT & \textbf{0.890} & 0.123 & \textbf{0.488} & 0.748 \\
 &  &  & Pythia-1.4B & 0.889 & 0.126 & 0.457 & \textbf{0.750} \\
 &  &  & Pythia-160m & 0.883 & \textbf{0.178} & 0.454 & 0.711 \\
\cline{3-8}
 &  & \multirow[t]{3}{*}{1} & Gemma-1.1-2B-IT & 0.880 & \textbf{0.186} & 0.495 & 0.750 \\
 &  &  & Pythia-1.4B & \textbf{0.889} & 0.183 & \textbf{0.502} & \textbf{0.764} \\
 &  &  & Pythia-160m & 0.873 & 0.182 & 0.454 & 0.738 \\
\cline{1-8}
 \multirow[t]{9}{*}{newsQA} & \multirow[t]{9}{*}{Gemma-1.1-2B-IT} & \multirow[t]{3}{*}{0.1} & Gemma-1.1-2B-IT & 0.874 & 0.166 & 0.413 & 0.687 \\
 &  &  & Pythia-1.4B & 0.871 & 0.294 & 0.505 & 0.642 \\
 &  &  & Pythia-160m & \textbf{0.902} & \textbf{0.482} & \textbf{0.530} & \textbf{0.800} \\
\cline{3-8}
 &  & \multirow[t]{3}{*}{0.2} & Gemma-1.1-2B-IT & \textbf{0.867} & 0.365 & 0.452 & 0.681 \\
 &  &  & Pythia-1.4B & 0.865 & 0.358 & \textbf{0.576} & 0.653 \\
 &  &  & Pythia-160m & 0.862 & \textbf{0.411} & 0.455 & \textbf{0.820} \\
\cline{3-8}
 &  & \multirow[t]{3}{*}{1} & Gemma-1.1-2B-IT & 0.889 & 0.148 & 0.443 & 0.742 \\
 &  &  & Pythia-1.4B & \textbf{0.890} & 0.222 & 0.519 & 0.723 \\
 &  &  & Pythia-160m & 0.862 & \textbf{0.411} & \textbf{0.455} & \textbf{0.821} \\
\cline{2-8}
 \multirow[t]{9}{*}{} & \multirow[t]{9}{*}{Pythia-1.4B} & \multirow[t]{3}{*}{0.1} & Gemma-1.1-2B-IT & \textbf{0.870} & 0.202 & \textbf{0.462} & 0.693 \\
 &  &  & Pythia-1.4B & 0.849 & 0.123 & 0.349 & 0.583 \\
 &  &  & Pythia-160m & 0.862 & \textbf{0.320} & \textbf{0.462} & \textbf{0.755} \\
\cline{3-8}
 &  & \multirow[t]{3}{*}{0.2} & Gemma-1.1-2B-IT & 0.887 & 0.163 & \textbf{0.504} & \textbf{0.778} \\
 &  &  & Pythia-1.4B & 0.876 & 0.168 & 0.468 & 0.713 \\
 &  &  & Pythia-160m & \textbf{0.892} & \textbf{0.352} & 0.412 & 0.735 \\
\cline{3-8}
 &  & \multirow[t]{3}{*}{1} & Gemma-1.1-2B-IT & \textbf{0.878} & 0.132 & \textbf{0.524} & \textbf{0.806} \\
 &  &  & Pythia-1.4B & 0.865 & 0.177 & 0.389 & 0.670 \\
 &  &  & Pythia-160m & 0.847 & \textbf{0.316} & 0.380 & 0.760 \\
  \bottomrule
 \end{tabular}
\end{table*}

\begin{table*}
\caption{\blue{ The table compares cross-validated attack metrics for $k=1$ with the attacker having a paraphrased version the text in the test queries for gemma-7B-IT. In each column, we highlight the metrics corresponding to the value of '$p$' where our attack exceeds the baselines.}}
    \centering
\label{tab:gamma7b}
\begin{tabular}{l|l|lrrrrr}
\toprule
model &   attack & $k$ & $p$   & AUC & TPR@1\%FPR & TPR@5\%FPR & TPR@10\%FPR \\
\midrule
\multirow[t]{26}{*}{SQuADShifts} & \multirow[t]{2}{*}{brainwash} & 1 & - & 0.529 & 0.011 & 0.056 & 0.112 \\
\cline{2-8} 
 & \multirow[t]{2}{*}{repeat} & 1 & - & 0.541 & 0.020 & 0.076 & 0.141 \\

\cline{2-8} 
 & \multirow[t]{2}{*}{logit} & 1 & - & 0.840 & 0.046 & 0.228 & 0.449 \\

\cline{2-8} 
 & \multirow[t]{6}{*}{reference (our)} & \multirow[t]{3}{*}{1} & 0.1 & \textbf{0.894} & \textbf{0.174} & \textbf{0.515} & \textbf{0.750} \\
 &  &  & 0.2 & 0.893 & 0.148 & 0.486 & 0.750 \\
 &  &  & 1 & 0.900 & 0.197 & 0.585 & 0.758 \\
\cline{2-8} 
 & \multirow[t]{14}{*}{prediction only (our)} & \multirow[t]{7}{*}{1} & 0.05 & 0.636 & 0.041 & 0.125 & 0.206 \\
 &  &  & 0.1 & 0.623 & 0.037 & 0.098 & 0.183 \\
 &  &  & 0.2 & 0.627 & 0.026 & 0.127 & 0.233 \\
 &  &  & 0.3 & 0.606 & 0.024 & 0.112 & 0.172 \\
 &  &  & 0.5 & 0.581 & 0.025 & 0.132 & 0.222 \\
 &  &  & 0.8 & 0.593 & 0.024 & 0.093 & 0.205 \\
 &  &  & 1 & 0.573 & 0.023 & 0.079 & 0.155 \\

\bottomrule
\multirow[t]{26}{*}{SQuAD} & \multirow[t]{2}{*}{brainwash} & 1 & - & 0.529 & 0.012 & 0.058 & 0.116 \\
\cline{2-8} 
 & \multirow[t]{2}{*}{repeat} & 1 & - & 0.546 & 0.026 & 0.068 & 0.120 \\
\cline{2-8} 
 & \multirow[t]{2}{*}{logit} & 1 & - & 0.587 & 0.037 & 0.092 & 0.155 \\
\cline{2-8}
 & \multirow[t]{6}{*}{reference (our)} & \multirow[t]{3}{*}{1} & 0.1 & \textbf{0.854} & \textbf{0.083} & \textbf{0.375} & \textbf{0.651} \\
 &  &  & 0.2 & 0.874 & 0.106 & 0.444 & 0.760 \\
 &  &  & 1 & 0.876 & 0.256 & 0.498 & 0.611 \\
\cline{2-8}
 & \multirow[t]{14}{*}{prediction only (our)} & \multirow[t]{7}{*}{1} & 0.05 & \textbf{0.687} & \textbf{0.067} & \textbf{0.176} & \textbf{0.286} \\
 &  &  & 0.1 & 0.727 & 0.115 & 0.183 & 0.309 \\
 &  &  & 0.2 & 0.674 & 0.054 & 0.217 & 0.283 \\
 &  &  & 0.3 & 0.631 & 0.053 & 0.142 & 0.210 \\
 &  &  & 0.5 & 0.596 & 0.038 & 0.125 & 0.187 \\
 &  &  & 0.8 & 0.594 & 0.078 & 0.102 & 0.160 \\
 &  &  & 1 & 0.563 & 0.020 & 0.034 & 0.112 \\

\bottomrule
\multirow[t]{26}{*}{newsQA} & \multirow[t]{2}{*}{brainwash} & 1 & - & 0.522 & 0.011 & 0.053 & 0.107 \\
\cline{2-8} 
 & \multirow[t]{2}{*}{repeat} & 1 & - & 0.572 & 0.012 & 0.052 & 0.155 \\
\cline{2-8} 
 & \multirow[t]{2}{*}{logit} & 1 & - & \textbf{0.929} & 0.373 & 0.604 & 0.744 \\
\cline{2-8} 
 & \multirow[t]{6}{*}{reference (our)} & \multirow[t]{3}{*}{1} & 0.1 & 0.886 & \textbf{0.593} & \textbf{0.617} & \textbf{0.648} \\
 &  &  & 0.2 & 0.807 & 0.441 & 0.455 & 0.473 \\
 &  &  & 1 & 0.896 & 0.639 & 0.653 & 0.669 \\
\cline{2-8} 
 & \multirow[t]{14}{*}{prediction only (our)} & \multirow[t]{7}{*}{1} & 0.05 & 0.559 & 0.036 & 0.080 & 0.178 \\
 &  &  & 0.1 & 0.532 & 0.020 & 0.097 & 0.151 \\
 &  &  & 0.2 & 0.536 & 0.016 & 0.085 & 0.137 \\
 &  &  & 0.3 & 0.581 & 0.015 & 0.064 & 0.150 \\
 &  &  & 0.5 & 0.568 & 0.800 & 0.044 & 0.102 \\
 &  &  & 0.8 & 0.541 & 0.025 & 0.128 & 0.177 \\
 &  &  & 1 & 0.537 & 0.037 & 0.093 & 0.168 \\
\bottomrule
\end{tabular}
\end{table*}

\vspace{3mm}

%% file: defenses.tex
\begin{table*}[htbp]    
\caption{\blue{The table compares cross-validated metrics for the second attack against 1-shot ICL for several values of $m$ when the ensemble prompting defense is applied.}}
\label{tab:defense}
\centering
\begin{tabular}{llclrrr}
\toprule
dataset &model & \blue{\makecell{number of \\ ensembles (m)}} & AUC & TPR@1\%FPR & TPR@5\%FPR & TPR@10\%FPR \\
\midrule
\multirow[t]{4}{*}{SQuAD} & Gemma-1.1-2B-IT & 2 &    0.524 & 0.016 & 0.051 & 0.132 \\
& & 3 &    0.550 & 0.030 & 0.084 & 0.162 \\
& & 4 &     0.539 & 0.043 & 0.093 & 0.169 \\
& & 5 &    0.570 & 0.044 & 0.090 & 0.152 \\
\midrule
\multirow[t]{4}{*}{} & Pythia-1.4B & 2 &   0.504 & 0.011 & 0.048 & 0.093 \\
& & 3 &   0.554 & 0.022 & 0.100 & 0.165 \\
& & 4 &   0.553 & 0.028 & 0.090 & 0.160  \\
& & 5 &   0.518 & 0.013 & 0.053 & 0.106 \\
 \midrule
SQuADShifts & \multirow[t]{4}{*}{Gemma-1.1-2B-IT} & 2 &  0.547 & 0.020 & 0.074 & 0.158 \\
& & 3  &  0.562 & 0.040 & 0.084 & 0.176 \\
& & 4  &  0.598 & 0.034 & 0.098 & 0.144 \\
& & 5  & 0.538 & 0.032 & 0.092 & 0.156 \\
\midrule
\multirow[t]{4}{*}{} & Pythia-1.4B & 2  & 0.551 & 0.086 & 0.152 & 0.203 \\
& & 3 &  0.580 & 0.120 & 0.162 & 0.265 \\
& & 4 &   0.546 & 0.134 & 0.150 & 0.270 \\
& & 5 &   0.559 & 0.068 & 0.098 & 0.215 \\
 \midrule
newsQA & \multirow[t]{4}{*}{Gemma-1.1-2B-IT} & 2 &  0.697 & 0.259 & 0.295 & 0.400 \\
& & 3  & 0.628 & 0.153 & 0.204 & 0.278 \\
& & 4  & 0.576 & 0.109 & 0.147 & 0.225 \\
& & 5  & 0.546 & 0.080 & 0.125 & 0.210 \\
\midrule
\multirow[t]{4}{*}{} & Pythia-1.4B & 2  & 0.551 & 0.086 & 0.152 & 0.203 \\
& & 3 &  0.580 & 0.120 & 0.162 & 0.265 \\
& & 4 &   0.546 & 0.134 & 0.150 & 0.270 \\
& & 5 &   0.559 & 0.068 & 0.098 & 0.215 \\
\bottomrule
\end{tabular}
\end{table*}

\section{Defenses}

\subsection{Administrative defenses}
A simple defense against the reference model attack is to accept the query components i.e. text and question separately, and combine those at the server's end using de-marketer words. This will ensure that the LM will predict only the answer tokens, and not the text tokens. We observed reduction in the success rates when the attacker trains the regression model only with the predicted answer words. This is because the answers are generally much shorter the text itself. The server can also prevent the maximum number of output tokens from being set to one. This can prevent generating highly granular predictions that could carry the strongest membership signals. 
\par To prevent the second attack, it is important to detect if the query text is complete or if it has been truncated, specifically if its suffix is missing. The server could employ an LLM-as-judge~\cite{llmasjudge} model that can perform structural and grammatical checks on the last sentence in the text against a checklist. For example, it can look for any missing terminal punctuations, obvious incomplete structures such as commas, parentheses, and brackets. An additional check is to compare the responses of zero-shot and few-shot ICL. If the application mandates that the text accompanying each question contains the information required to answer it, and the zero-shot response indicates missing information while the few-shot response is informative, this suggests the few-shot answer is likely influenced by the in-context examples. The structural compliance requirements for the query text could be stricter in such cases.

\subsection{Methodological defenses}
The work \cite{iclmia3} proposed ensemble prompting as a defense against MIAs for classification tasks. They query the LM with different prompts, and the prediction probabilities for all possible target classes are aggregated and used for the final prediction. 
\blue{
We extend this method to the DQA task: Let $m$ denote the number of ensembles. The server retrieves $k \times m$ NNs for a query, prompting the LM $m$ times, each with a $k$ in-context examples. It then computes the $c$ most frequent words from the $m$ answers and re-prompts the LM to ask for a final answer based on these frequent words, which is then served to the user. The ensembling strategy reduces the impact of a single query when it appears as an in‑context demonstration; only one prompt contains that query while the others feature different examples, thereby weakening the membership cues. To verify this defense, we consider the case when the texts in the test queries and member index are identical. This is the ideal situation from the adversary's point of view. For $m \geq 2$, Tables~\ref{tab:defense} and \ref{tab:defense_gemma7b} (Appendix) show the revised success metrics for our attacks when the ensemble prompting defense is applied. The attacks are conducted using all prefixes i.e. with $p=1$. In most cases, TPR@low FPR values drop by up to 5 fold, when compared to Tables~\ref{tab:shifted_squad_exact}, \ref{tab:squad_exact} and \ref{tab:newsqa_exact}.
}

\subsection{Discussion on potential DP defenses}
The framework of differential privacy (DP) provides mathematically rigorous and robust protection against membership inference attacks, making DP-ICL an active area of research~\citep{tangprivacy,gao2025data,amin2024private,wuprivacy}. Ensembling is commonly used in DP solutions to convert a prediction problem into an aggregation problem. Among available solutions, the DP-KSA method from \cite{wuprivacy} privatizes the released most frequent words obtained from the procedure mentioned above. Alternatively, methods from \cite{amin2024private,wuprivacy} generate the answer
texts token-by-token by privately releasing the mean logits from ensemble prompts and by choosing the token with the largest weight as the next token. Adapting the existing DP solutions as they are to our case might lead to harsh privacy-utility trade-offs. This is because, unlike our setting, these solutions select in-context examples via Poisson sampling to amplify the privacy guarantees. In contrast, retrieving kNNs is a deterministic process. The privacy loss without any amplification mechanism can blowup quickly even after a small number of DP compositions. Therefore developing a practical ensemble prompting based DP solution for retrieval powered ICL is non-trivial and justifies a new work. Specifically, a new solution needs to be designed and analyzed for the two opposing effects at play: (a) enhanced utility from more relevant demonstrations, and (b) a degraded privacy‑utility trade-off due to absence of a privacy amplification mechanism. The recent work from~\cite{dpknnicl} takes a step in this direction.


%% file: limitations.tex
\section{Conclusion}
\subsection{Final Remarks}
We propose two prefix based attacks for membership inference against retrieval based ICL used for DQA tasks. As per our knowledge, retrieval augmented ICL has not been attacked before our work despite its practical relevance. Unlike baselines, our attacks do not aim to predict large number of tokens in a single API call or risk overflowing the LM's context-window. To maintain stealth, we can execute these attacks with any order of prefixes and over time. Our evaluations with paraphrased queries show that we can achieve strong true positive rates with relaxed assumptions. Our work highlights the need of developing a DP solution for this problem with a favorable privacy-utility tradeoff.

\subsection{Limitations}
To balance experiment coverage and cost, our experiments are limited to 1‑shot ICL and employ greedy decoding. For sampling based decoding methods, we acknowledge that the variance introduced into the token‑selection process could affect the attack performance.
\par While our attacks predict a small number of tokens in each API call to maintain stealth, and we can successfully infer membership with a small subset of prefixes, the number of API calls required for each membership inference still scales with the query length. 

\subsection{Future Work}
Finding clever strategies to reduce the number of queries while retaining stealth and strong discriminating ability is an important future direction. Other directions include attacking ICL solutions that utilize alternative sample‑selection methods, such as those based on perplexity scores~\cite{ppl_based} or reinforcement learning~\cite{rl_icl}. It is also appealing to extend this attack to other retrieval systems. 
Finally, an interesting research avenue is to design mechanisms that combine retrieval with DP-based aggregation, thereby ensuring the formal privacy guarantees of DP while mitigating the utility degradation brought by random sampling of demonstrations.

\subsection{Acknowledgment}
The authors are grateful to the anonymous reviewers for their insightful comments, which helped improve this work.

%% file: baselines.tex
\begin{algorithm*}
\caption{Logit, inspired from \cite{iclmia1,iclmia3}}\label{alg:logit}
\begin{algorithmic}[1]

\Function{Logit}{$\mathcal{Q}$}
\State $\langle t,q,? \rangle = \mathcal{Q}$
 \State $Q$ = 'text:' + $t$ + 'question:' + $q$  +'answer:'
\State $P^{v} = \mathrm{LM}^{v} \big(\mathrm{BuildPrompt}(D,Q,t,k)\big)$
\State $L^{v} \in \mathbb{R}^{m \times V}$ be the logits obtained from the previous LM call. 
\State T be the list of token ids of the ground truth answer a.
\State $\ell = 0$ 

\For{ $i \in $\big[min$(m,|T|)\big]$} \Comment{Compute the mean of logits assigned to the tokens of the answer a.}

\State $\ell = \ell + L^{v}_{i}[T_i]$
\EndFor
\State \Return $\frac{\ell}{\mathrm{min}(m,|T|)}$ \Comment{We compute the mean logit score for the true answer.}

    \EndFunction

\end{algorithmic}
\end{algorithm*}
\begin{algorithm*}
\caption{Repeat, inspired from \cite{iclmia2}}\label{alg:repeat}
\begin{algorithmic}[1]

\Function{Repeat}{$\mathcal{Q}$, $p$}
\State $\langle t,q,? \rangle = \mathcal{Q}$
\State  $i = p \times len(t)$
 \State   $m =len(t)-i$
 \State Q = 'text:' + $t_{:i}$ 
\State $P^{v} = \mathrm{LM} \big(\mathrm{BuildPrompt}(D,Q,t_{:i},k),\maxtokens=m\big)$ \Comment{$LM$ will try to predict the suffix $t_{i+1:}$ of the query text.}
\State $P'$ be the prefix of $\mathrm{P}^{v}$ that is a prediction the text suffix $t_{i+1:}$. 
\State \Return  $\ell=f(P',t_{i+1:})$ be the embedding distance between $P'$ and $t_{i+1:}$.
    \EndFunction

\end{algorithmic}
\end{algorithm*}

\begin{algorithm*}
\caption{Brainwash, inspired from \cite{iclmia2}}\label{alg:brainwash}
\begin{algorithmic}[1]

\Function{Brainwash}{$\mathcal{Q}$, $T$}
\State $\langle t,q,? \rangle = \mathcal{Q}$
\State \dummy = $\perp$
 \State $Q$ = 'text:' + $t$ + 'question:' + $q$  +'answer:' 

\For{ $i \in [T]$}
\Comment{Iteratively prepend copies of $\mathcal{Q}$ with \dummy to brainwash LM to respond with a wrong answer.}
\State $Q$ = 'text:' + $t$ + 'question:' + $q$  +'answer:' + \dummy + $Q$
\State $P^{v} = \mathrm{LM}\big(\mathrm{BuildPrompt}(D,Q,t,k)\big)$
\If{$P^v = \dummy$}
\State \Return $i$
\EndIf
\EndFor

\State \Return $i$ \Comment{We use the iteration at which the LM replied with \dummy as the score.}

    \EndFunction

\end{algorithmic}
\end{algorithm*}

%% file: appendix.tex
\begin{algorithm}[htbp]
\caption{Prompt builder For document question answering.}\label{alg:promptQA}
\begin{algorithmic}

\Function{$\mathrm{BuildPrompt}$}{$D,Q,t,k$}
\State  Let  $E=R_k(D,t)$ be the list of retrieved kNN's serving as in-context examples. 
\State $e$ = ''
\For{ $\langle t,q,a \rangle \in E$}
 \State $e$  += 'text:' + $t$ + 'question:' + $q$  +'answer:' + $a$
\EndFor
\State \Return prompt = $I + e + Q$
\EndFunction
\end{algorithmic}
\end{algorithm}

\begin{table*}
\caption{\textbf{SQuADShifts}: The table compares cross-validated attack metrics, assuming that the paragraphs in the attacker's test queries and service provider's demonstration dataset are identical. In comparison, the texts in the attacker's queries are paraphrased for the results in Figure~\ref{tab:shifted_squad_para}.}
\label{tab:shifted_squad_exact}
\centering
\begin{tabular}{ll|r|r|r|r|r|r}
\toprule
 &  &  &  & AUC & TPR@1\%FPR & TPR@5\%FPR & TPR@10\%FPR \\
model & $k$ & attack & $p$ &  &  &  &  \\
\midrule
\multirow[t]{13}{*}{Gemma-1.1-2B-IT} & \multirow[t]{13}{*}{1} & brainwash & - & 0.834 & 0.166 & 0.465 & 0.559 \\
\cline{3-8}
 &  & repeat & - & 0.779 & 0.134 & 0.462 & 0.534 \\
\cline{3-8}
 &  & logit & - & 0.922 & 0.160 & 0.615 & 0.844 \\
\cline{3-8}
 &  & \multirow[t]{3}{*}{reference (our)} & 0.1 & 0.889 & 0.148 & 0.496 & 0.732 \\
 &  &  & 0.2 & 0.879 & 0.153 & 0.499 & 0.743 \\
 &  &  & 1 & 0.900 & \textbf{0.183} & 0.529 & 0.772 \\
\cline{2-8}
 &  & \multirow[t]{7}{*}{prediction only (our)} & 0.05 & 0.807 & 0.083 & 0.319 & 0.502 \\
 &  &  & 0.1 & 0.849 & 0.105 & 0.360 & 0.542 \\
 &  &  & 0.2 & 0.847 & 0.057 & 0.316 & 0.557 \\
 &  &  & 0.3 & 0.861 & 0.124 & 0.381 & 0.618 \\
 &  &  & 0.5 & 0.858 & 0.092 & 0.413 & 0.675 \\
 &  &  & 0.8 & 0.875 & 0.109 & 0.459 & 0.692 \\
 &  &  & 1 & 0.867 & 0.108 & 0.393 & 0.679 \\
\midrule
\multirow[t]{13}{*}{pythia-1.4b} & \multirow[t]{13}{*}{1} & brainwash & - & 0.508 & 0.018 & 0.058 & 0.108 \\
\cline{3-8}
 &  & repeat & - & 0.863 & 0.664 & 0.686 & 0.716 \\
\cline{3-8}
 &  & logit & - & 0.988 & 0.678 & 0.990 & 0.994 \\
\cline{3-8}
 &  & \multirow[t]{3}{*}{reference (our)} & 0.1 & 0.862 & 0.132 & 0.428 & 0.658 \\
 &  &  & 0.2 & 0.877 & 0.117 & 0.482 & 0.681 \\
 &  &  & 1 & 0.883 & 0.136 & 0.493 & 0.712 \\
\cline{3-8}
 &  & \multirow[t]{7}{*}{prediction only (our)} & 0.05 & 0.832 & 0.098 & 0.372 & 0.549 \\
 &  &  & 0.1 & 0.846 & 0.058 & 0.300 & 0.582 \\
 &  &  & 0.2 & 0.870 & 0.084 & 0.419 & 0.674 \\
 &  &  & 0.3 & 0.866 & 0.077 & 0.348 & 0.599 \\
 &  &  & 0.5 & 0.877 & 0.129 & 0.477 & 0.722 \\
 &  &  & 0.8 & 0.868 & 0.097 & 0.360 & 0.641 \\
 &  &  & 1 & 0.869 & 0.089 & 0.372 & 0.646 \\
 \bottomrule
\end{tabular}
\end{table*}

\begin{table*}
\caption{\textbf{SQuAD}: The table compares cross-validated attack metrics, assuming that the paragraphs in the attacker's test queries and service provider's demonstration dataset are identical. In comparison, the texts in the attacker's queries are paraphrased for the results in Figure~\ref{tab:squad_para}.}
\label{tab:squad_exact}
\centering
\begin{tabular}{ll|r|r|r|r|r|r}
\toprule
model & $k$ & attack & $p$ & AUC & TPR@1\%FPR & TPR@5\%FPR & TPR@10\%FPR \\
\midrule
\multirow[t]{13}{*}{gemma-1.1-2b-it} & \multirow[t]{13}{*}{1} & brainwash & - & 0.808 & 0.141 & 0.378 & 0.499 \\
\cline{3-8}
 &  & repeat & - & 0.761 & 0.114 & 0.379 & 0.520 \\
\cline{3-8}
 &  & logit & - & 0.873 & 0.079 & 0.351 & 0.654 \\
\cline{3-8}
 &  & \multirow[t]{3}{*}{reference (our)} & 0.1 & \textbf{0.850} & 0.112 & 0.379 & 0.598 \\
 &  &  & 0.2 & 0.869 & \textbf{0.158} & \textbf{0.430} & \textbf{0.707} \\
 &  &  & 1 & 0.900 & 0.286 & 0.659 & 0.792 \\
\cline{3-8}
 &  & \multirow[t]{7}{*}{prediction only (our)} & 0.05 & 0.854 & 0.093 & 0.383 & 0.592 \\
 &  &  & 0.1 & 0.869 & 0.164 & \textbf{0.474} & \textbf{0.670} \\
 &  &  & 0.2 & 0.874 & 0.079 & 0.385 & 0.649 \\
 &  &  & 0.3 & 0.874 & 0.089 & 0.443 & 0.671 \\
 &  &  & 0.5 & 0.889 & 0.143 & 0.439 & 0.670 \\
 &  &  & 0.8 & 0.864 & 0.066 & 0.332 & 0.624 \\
 &  &  & 1 & 0.873 & 0.100 & 0.389 & 0.653 \\
\midrule
\multirow[t]{13}{*}{pythia-1.4b} & \multirow[t]{13}{*}{1} & brainwash & - & 0.503 & 0.015 & 0.056 & 0.105 \\
\cline{3-8}
 &  & repeat & - & 0.832 & 0.616 & 0.654 & 0.686 \\
\cline{3-8}
 &  & logit & - & 0.900 & 0.115 & 0.564 & 0.825 \\
\cline{3-8}
 &  & \multirow[t]{3}{*}{reference (our)} & 0.1 & 0.867 & 0.152 & 0.422 & 0.641 \\
 &  &  & 0.2 & 0.883 & 0.121 & 0.438 & 0.656 \\
 &  &  & 1 & 0.880 & 0.175 & 0.504 & 0.740 \\
\cline{3-8}
 &  & \multirow[t]{7}{*}{prediction only (our)} & 0.05 & 0.856 & 0.082 & 0.330 & 0.657 \\
 &  &  & 0.1 & 0.869 & 0.144 & 0.417 & 0.690 \\
 &  &  & 0.2 & 0.881 & 0.091 & 0.475 & 0.730 \\
 &  &  & 0.3 & 0.877 & 0.079 & 0.395 & 0.704 \\
 &  &  & 0.5 & 0.882 & 0.163 & 0.506 & 0.725 \\
 &  &  & 0.8 & 0.888 & 0.109 & 0.365 & 0.762 \\
 &  &  & 1 & 0.875 & 0.122 & 0.451 & 0.715 \\
\bottomrule
\end{tabular}
\end{table*}

\begin{table*}
\caption{\textbf{newsQA}: The table compares cross-validated attack metrics, assuming that the paragraphs in the attacker's test queries and service provider's demonstration dataset are identical. In comparison, the texts in the attacker's queries are paraphrased for the results in Figure~\ref{tab:newsqa_para}.}
    \centering
\label{tab:newsqa_exact}
    \begin{tabular}{ll|r|r|r|r|r|r}
\toprule
model & $k$ & attack & $p$   & AUC & TPR@1\%FPR & TPR@5\%FPR & TPR@10\%FPR \\
\midrule
\multirow[t]{13}{*}{gemma-1.1-2b-it} & \multirow[t]{13}{*}{1} & brainwash & - & 0.779 & 0.145 & 0.333 & 0.454 \\
\cline{3-8}
 &  & repeat & - & 0.797 & 0.060 & 0.280 & 0.478 \\
\cline{3-8}
 &  & logit & - & 0.950 & 0.370 & 0.817 & 0.937 \\
\cline{3-8}
 &  & \multirow[t]{3}{*}{reference (our)} & 0.1 & 0.874 & 0.104 & 0.466 & 0.668 \\
 &  &  & 0.2 & 0.892 & 0.192 & 0.481 & 0.720 \\
 &  &  & 1 & 0.876 & 0.107 & 0.431 & 0.677 \\
\cline{3-8}
 &  & \multirow[t]{7}{*}{prediction only (our)} & 0.05 & 0.874 & 0.132 & 0.596 & 0.701 \\
 &  &  & 0.1 & 0.872 & 0.179 & 0.545 & 0.700 \\
 &  &  & 0.2 & 0.901 & 0.263 & 0.636 & 0.756 \\
 &  &  & 0.3 & 0.866 & 0.192 & 0.574 & 0.692 \\
 &  &  & 0.5 & 0.890 & 0.175 & 0.604 & 0.715 \\
 &  &  & 0.8 & 0.899 & 0.429 & 0.682 & 0.828 \\
 &  &  & 1 & 0.894 & 0.241 & 0.661 & 0.837 \\
 \midrule
 \multirow[t]{13}{*}{pythia-1.4b} & \multirow[t]{13}{*}{1} & brainwash & - & 0.636 & 0.101 & 0.267 & 0.354 \\
\cline{3-8}
 &  & repeat & - & 0.979 & 0.925 & 0.962 & 0.962 \\
\cline{3-8}
 &  & logit & - & 0.993 & 0.891 & 0.982 & 0.992 \\
\cline{3-8}
 &  & \multirow[t]{3}{*}{reference (our)} & 0.1 & 0.864 & 0.119 & 0.374 & 0.638 \\
 &  &  & 0.2 & 0.886 & 0.167 & 0.489 & 0.678 \\
 &  &  & 1 & 0.891 & 0.148 & 0.486 & 0.695 \\
\cline{3-8}
 &  & \multirow[t]{7}{*}{prediction only (our)} & 0.05 & 0.915 & 0.480 & 0.641 & 0.741 \\
 &  &  & 0.1 & 0.882 & 0.380 & 0.598 & 0.719 \\
 &  &  & 0.2 & 0.871 & 0.230 & 0.411 & 0.677 \\
 &  &  & 0.3 & 0.859 & 0.190 & 0.414 & 0.573 \\
 &  &  & 0.5 & 0.899 & 0.182 & 0.643 & 0.720 \\
 &  &  & 0.8 & 0.888 & 0.199 & 0.443 & 0.751 \\
 &  &  & 1 & 0.877 & 0.079 & 0.512 & 0.701 \\
\bottomrule
\end{tabular}

\end{table*}

\begin{table*}[htbp] 
\blue{
\caption{The table compares cross-validated metrics for the second attack against 1-shot ICL on gemma-7B-IT for several values of $m$ when the ensemble prompting defense is applied.}}
\label{tab:defense_gemma7b}
\centering
\begin{tabular}{llclrrr}
\toprule
dataset &model & \makecell{number of \\ ensembles (m)} & AUC & TPR@1\%FPR & TPR@5\%FPR & TPR@10\%FPR \\
\midrule
\multirow[t]{4}{*}{SQuAD} & gemma-7B-IT & 2 &    0.524 & 0.013 & 0.053 & 0.112 \\
& & 3 &    0.530 & 0.031 & 0.074 & 0.142 \\
& & 4 &     0.544 & 0.022 & 0.053 & 0.139 \\
& & 5 &    0.542 & 0.023 & 0.060 & 0.122 \\
\midrule
SQuADShifts & \multirow[t]{4}{*}{gemma-7B-IT} & 2 &  0.547 & 0.021 & 0.084 & 0.138 \\
& & 3  &  0.522 & 0.041 & 0.064 & 0.116 \\
& & 4  &  0.518 & 0.024 & 0.058 & 0.114 \\
& & 5  & 0.496 & 0.012 & 0.072 & 0.116 \\
\midrule
newsQA & \multirow[t]{4}{*}{gemma-7B-IT} & 2 &  0.497 & 0.059 & 0.150 & 0.110 \\
& & 3  & 0.512 & 0.061 & 0.134 & 0.158 \\
& & 4  & 0.546 & 0.082 & 0.127 & 0.185 \\
& & 5  & 0.526 & 0.087 & 0.115 & 0.190 \\
\bottomrule
\end{tabular}
\end{table*}

\begin{figure*} [h!]
\caption{The distributions of mean negative log-likelihoods are presented for various reference models (X axes), with Pythia and Gemma used as the victim models. The labels pred\_1, pred\_0, true\_1, and true\_0 are for the boxplots of true and predicted values  for member (1) and non-member (0) queries. The true values are obtained from the target model.}
     \label{fig:qa_nll}
     \centering
     \subfigure[]{
        \includegraphics[width=.48\textwidth]{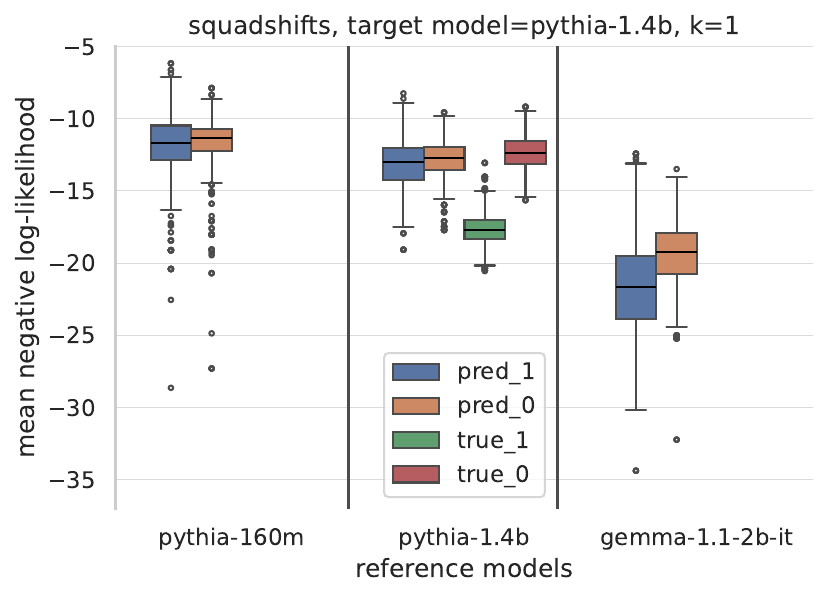}
        \label{fig:squadshift_pythia_nll}
     }
     \subfigure[]{
        \includegraphics[width=.48\textwidth]{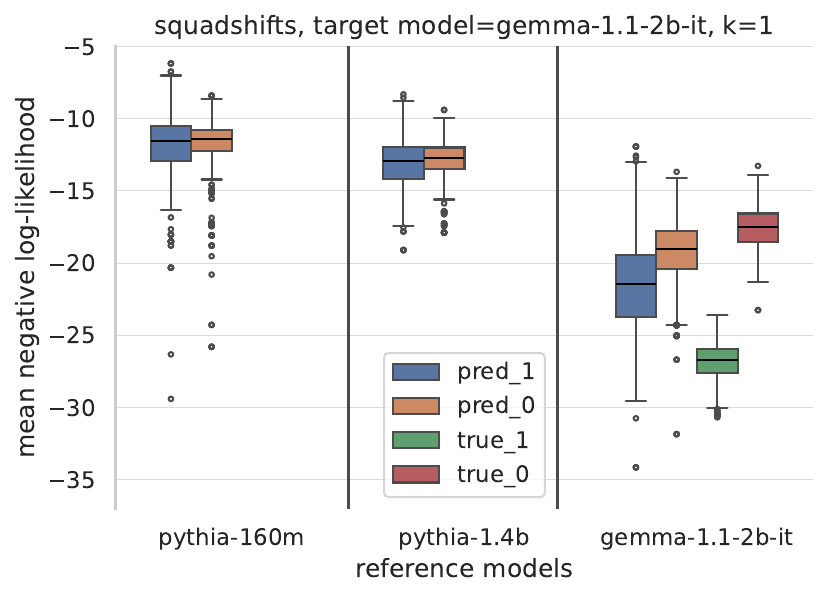}
        \label{fig:squadshift_gemma_nll}
     }
     \subfigure[]{
        \includegraphics[width=.48\textwidth]{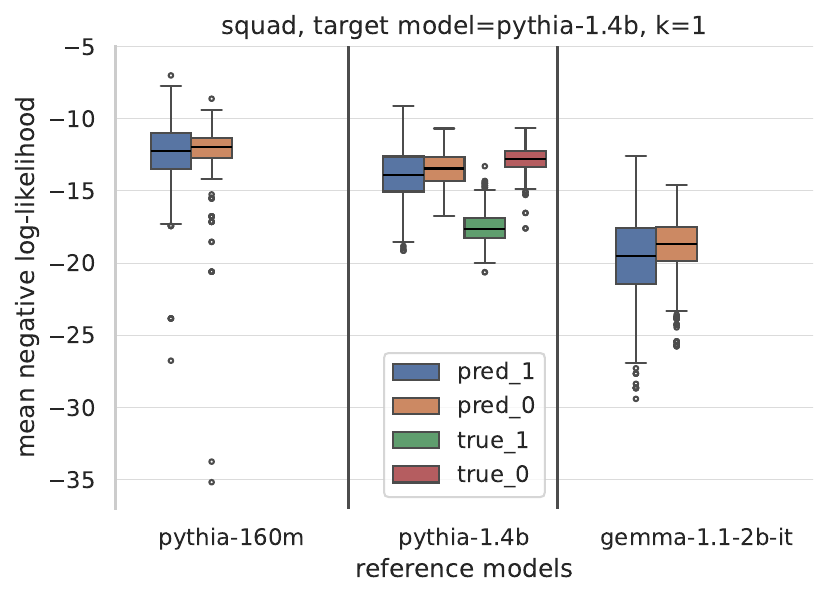}
        \label{fig:squad_pythia_nll}
     }
     \subfigure[]{
        \includegraphics[width=.48\textwidth]{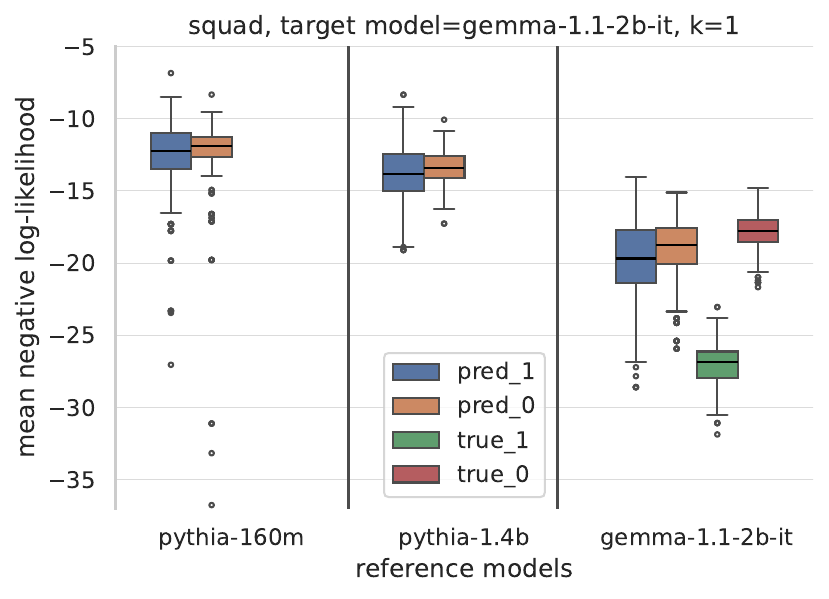}
        \label{fig:squad_gemma_nll}
     }
     \subfigure[]{
        \includegraphics[width=.48\textwidth]{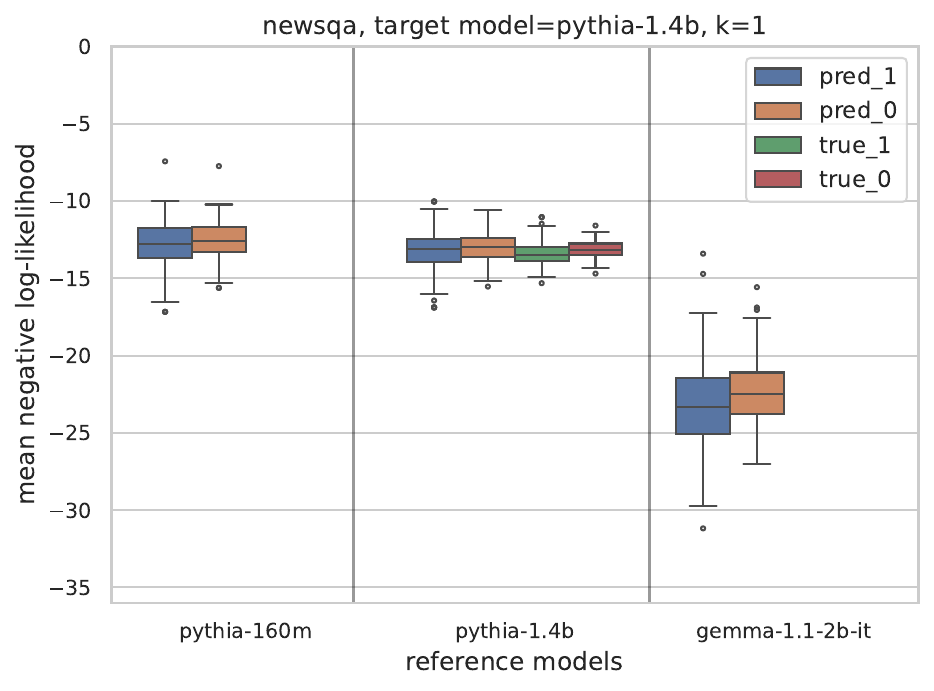}
        \label{fig:newsqa_pythia_nll}
     }
     \subfigure[]{
        \includegraphics[width=.48\textwidth]{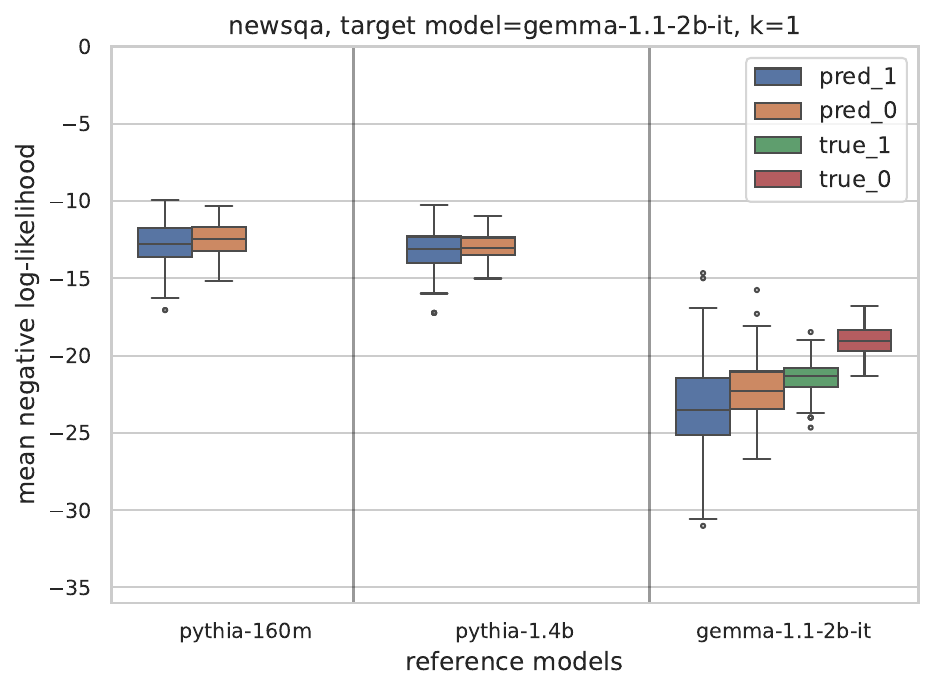}
        \label{fig:newsqa_gemma_nll}
     }
     
\end{figure*}

\subsection{Example DQA Prompt template}
\label{sec:qa_prompt}

For question answering task,we used the following prompt template. The bracketed terms (e.g., \texttt{\{demo text1\}}) indicate placeholders for specific data.
\begin{verbatim}
Read the text: 
{demo text1}
Answer the question with at most 10 words: 
{demo question1}
Do not provide a Yes/No answer: 
{demo answer1}

Read the text: 
{demo text2}
Answer the question with at most 10 words: 
{demo question2}
Do not provide a Yes/No answer: 
{demo Answer2}

Read the text: 
{query text}
Answer the question with at most 10 words: 
{query question}
Do not provide a Yes/No answer: 
\end{verbatim}

\clearpage
\section{Rebuttal}
\begin{figure*}\caption{\blue{Flow diagram for the attack from Algorithm~\ref{alg:ref_model}.} }

    \centering
    \includegraphics[width=1\linewidth]{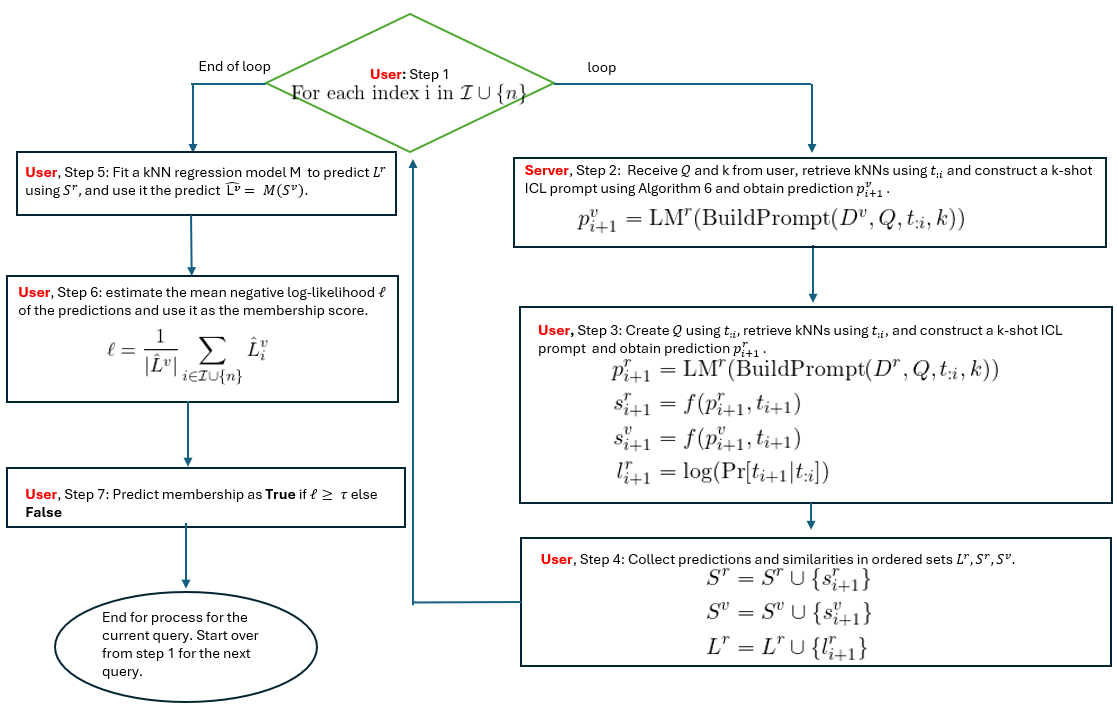}
    \label{fig:attack1}
\end{figure*}

\begin{figure*}
    \centering \caption{\blue{Flow diagram for the attack from Algorithm~\ref{alg:labelonly}.} }
    \includegraphics[width=1\linewidth]{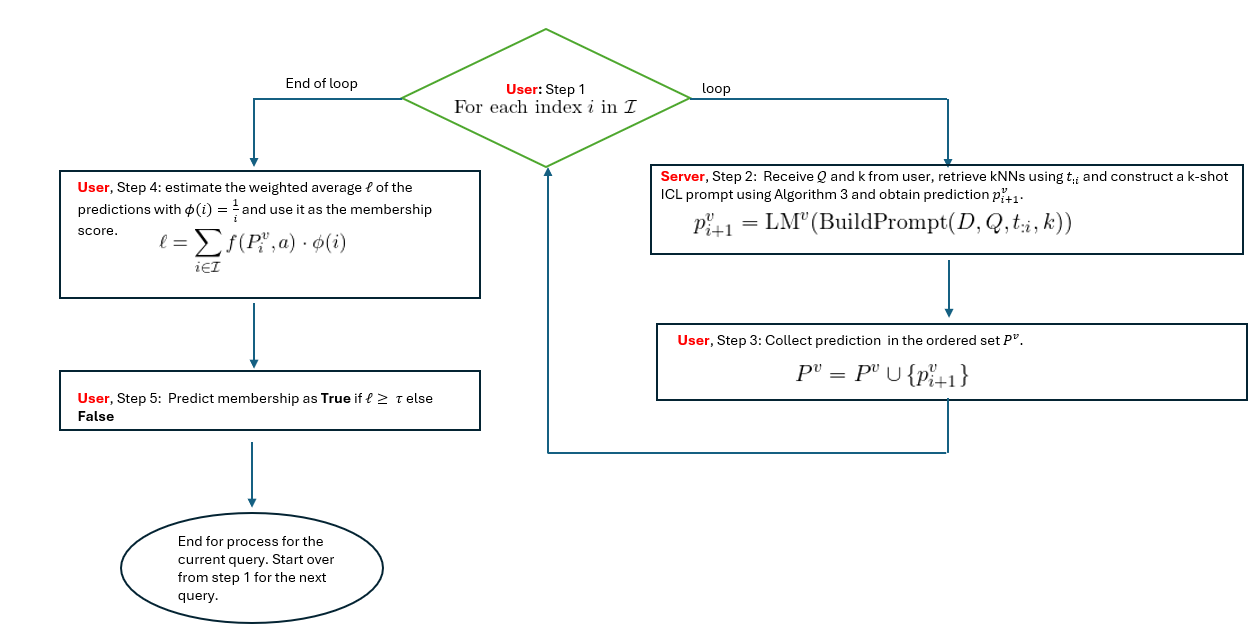}
    
    \label{fig:attack2}
\end{figure*}

